\documentclass[aps,prx,twocolumn,superscriptaddress,floatfix,noeprint]{revtex4-2}
\usepackage[utf8]{inputenc}
\usepackage[T1]{fontenc}
\usepackage{amssymb}
\usepackage{graphics}
\usepackage{float}
\graphicspath{{figures/}}
\usepackage{mathtools}
\usepackage{braket}
\usepackage{microtype}
\usepackage{bm}
\usepackage{enumitem}
\usepackage[colorlinks=true,allcolors=blue]{hyperref}

\DeclarePairedDelimiter\abs{\lvert}{\rvert}
\DeclarePairedDelimiter\mean{\langle}{\rangle}
\DeclarePairedDelimiter\norm{\lVert}{\rVert}

\DeclareMathOperator{\sign}{sign}
\DeclareMathOperator{\shortsin}{s}
\DeclareMathOperator{\shortcos}{c}
\DeclareMathOperator{\diag}{diag}

\DeclareMathOperator{\imag}{Im}
\DeclareMathOperator{\res}{Res}
\DeclareMathOperator{\K}{K}
\DeclareMathOperator{\J}{J}

\newcommand{\EBS}{E_\mathrm{BS}}

\newtheorem{theorem}{Theorem}

\begin{document}

% \title{Effective Models for Quantum Emitters Coupled to 2D Topological Insulators}
\title{Topological Effects in Two-Dimensional Quantum Emitter Systems}

\author{Miguel Bello}
\email{miguel.bello@mpq.mpg.de}
\affiliation{Max-Planck-Institut f\"ur Quantenoptik, Hans-Kopfermann-Stra{\ss}e 1, D-85748 Garching, Germany}

\author{J. I. Cirac}
\affiliation{Max-Planck-Institut f\"ur Quantenoptik, Hans-Kopfermann-Stra{\ss}e 1, D-85748 Garching, Germany}
\affiliation{Munich Center for Quantum Science and Technology, Schellingstra{\ss}e 4, 80799 M\"unchen, Germany}

\date{\today}

\begin{abstract}
    In this work we discuss particular effects that take place in systems of quantum emitters coupled to two-dimensional bosonic topological insulators.
    For a single emitter coupled to the Haldane model, we find a ``fragile'' quasibound state that makes the emitter dynamics very sensitive to the model's parameters, and gives rise to effective long-range interactions that break time-reversal symmetry.
    We then discuss one-dimensional arrangements of emitters, emitter line defects, and how the topology of the bath affects the effective polariton models that appear in the weak-coupling regime when the emitters are spectrally tuned to a bandgap.
    In the Harper-Hofstadter model we link the non-monotonic character of the effective interactions to the Chern numbers of the surrounding energy bands, while in the Haldane model we show that the effective models are either gapless or not depending on the topology of the bath. 
    Last, we discuss how the presence of emitters forming an ordered array, an emitter superlattice, can produce polariton models with non-trivial Chern numbers, and also modify the topology of the photonic states in the bath.
\end{abstract}

% \begin{abstract}
%     Topological properties of 1D photonic baths can be observed looking at the vacancy-like bound states that emerge when a single quantum emitter is coupled to the bath. In this work we show that the analogous phenomenon in 2D topological-insulator baths is to find fully directional bound states, and gapless effective single-particle models for 1D arrangements of emitters, which may support chiral currents. We also discuss the features of 2D arrangements (quantum-emitter superlattices), in particular, how the topology of the resulting polariton bands depends on the geometry of the arrangement and the light-matter coupling strength.  
% \end{abstract}

\maketitle

\section{Introduction \label{sec:intro}}

Quantum emitters coupled to structured baths display a number of phenomena that are not present in more conventional baths: bound states, non-Markovian dynamics, etc. \cite{GonzalezTudela2017PRL,GonzalezTudela2017PRA,Roos2020}
In particular, topological baths offer the possibility to go beyond the standard quantum electrodynamics of atoms in vacuum, and explore the interplay between exotic states of light and matter. 
Several theoretical studies \cite{Bello2019,Leonforte2021,Bernardis2021,Vega2021,Zhang2022,Vega2022,Roccati2022,Gong2022PRA,Gong2022PRL} and a few experimental works \cite{Barik2018,Kim2021} have already demonstrated interesting effects, such as the generation of exotic coherent dipole-dipole interactions \cite{Bello2019,Kim2021,Roccati2022}, or coupling to chiral edge states which are topologically protected \cite{Barik2018,Zhang2022,Vega2022}.
These works trigger further questions, as there are many different kinds of topological models that remain unexplored, and many different topological concepts whose implications for quantum optics are not yet well understood.

In one-dimensional (1D) topological photonic baths, the relation between the underlying bath topology and the quantum optical effects that can be observed is fairly well understood. 
For example, it has been shown that the bath topology plays a role in the bound states that the emitters seed. 
For specific values of the emitter transition frequency, the atoms can behave like a mirror or, in other words, as if they produce a vacancy in the bath. 
Consequently, the bound states can share many properties of the topological edge states that the bath supports in the nontrivial phase \cite{Bello2019,Vega2021}.
These vacancy-like bound states were later recognized as a general feature that occurs in many photonic lattices, whether topological or not \cite{Leonforte2021}.

In two-dimensional (2D) topological baths, the effects of a nontrivial topology are much less clear.
A straightforward way to observe topological effects is to couple emitters to the boundaries of 2D topological baths.
In this case, the dynamics are sensitive to whether or not the bath supports edge states, and this can be exploited for the development of interesting applications \cite{Zhang2022,Vega2022}.
Another possibility is to couple emitters to the bulk of a 2D topological bath, however, in this case it is hard to disentangle what effects are simply due to the breaking of time-reversal symmetry, and what effects are due to a nontrivial topology.
This kind of systems have been explored in Ref.~\cite{Bernardis2021} for a few emitters coupled to the Harper-Hofstadter model in the Landau regime, and in Ref.~\cite{Leonforte2021} for a single emitter coupled to the Haldane model. 
In the latter, it was shown that vacancy-like bound states exist only when the bath is in a nontrivial topological phase, albeit only for specific values of the parameters within these phases, where the model recovers particle-hole symmetry.
It is therefore an open question to what extent the topology of the underlying bath can affect the dynamics of quantum emitters coupled to it.

In this paper we aim to shed light on this issue.
We consider a 2D lattice of bosonic degrees of freedom that is a topological insulator, and a set of emitters coupled to it.
We investigate the dissipative and coherent dynamics of emitters, paying especial emphasis on the way that topological properties are reflected in the properties of polaritons (excitations trapped in atom-photon bound states).
We consider three different scenarios: i) few emitters; ii) a 1D array of emitters coupled along one of the columns (or rows) of the 2D lattice; iii) 2D arrays, where each emitter is coupled to a single bath mode in a periodic fashion. 
In case i), the physics is similar to that taking place in other topologically trivial lattices, albeit with some differences, such as the appearance of quasi-bound states around singular gaps which are not vacancy-like, and are therefore, in a sense, fragile, as they require finely tuned system parameters to exist.
In case ii), the bath's topology has more dramatic effects. The effective polariton models have dispersion relations whose complexity, or in which the presence of a gap, is linked to the Chern numbers of the bath's energy bands. 
This has clear observable consequences, such as propagation with multiple wavefronts, or the chiral dynamics of a single polariton in a 1D emitter array.
In case iii), we show how the polariton models can have different Chern numbers depending on the array geometry and the light-matter coupling strength.

This paper is organized as follows. 
In Section \ref{sec:models} we describe the specific models studied, and in Section \ref{sec:vbs} we recall the concept of vacancy-like bound states, which is essential to understand the following sections. 
Then, in Sections \ref{sec:few_emitters} to \ref{sec:emitter_superlattices} we describe the phenomena that take place in different kinds of emitter arrangements, in order of increasing complexity. 
We conclude in Section \ref{sec:conclusion} by summarizing the main results and pointing out possible future research directions.

\section{Models \label{sec:models}}

We consider a set of $N$ identical two-level systems (quantum emitters), with raising and lowering operators $\{\sigma^+_n \equiv \ket{e_n}\!\bra{g_n}, \sigma^-_n \equiv (\sigma^+_n)^\dag\}_{n=1}^N$; $\ket{g_n}$ and $\ket{e_n}$ denote the ground and excited state of the $n$th emitter, respectively. 
Each one is coupled to a single local bosonic mode of a 2D lattice of size $L_x \times L_y$ in units of the lattice constant, which is modelled via a tight-binding Hamiltonian $H_B$. 
The coupling, described by $H_I$, is of the Jaynes-Cummings type, with coupling constant $g$. 
% (hereafter referred to as the light-matter coupling constant). 
The total Hamiltonian of the system is $H = H_E + H_B + H_I$, with
\begin{align}
    H_E & = \Delta \sum_n \sigma^+_n\sigma^-_n \,, \\
    H_B & = \sum_{\bm{k}} \Psi^\dag_{\bm{k}} \mathcal{H}_B(\bm{k}) \Psi_{\bm{k}} \,, \\
    H_I & = \frac{g}{\sqrt{L_x L_y}} \sum_{n,\,\bm{k}} e^{i\bm{k}\bm{r}_n} \sigma^+_n a_{\alpha_n,\bm{k}} + \mathrm{H.c.}
\end{align}
$\Psi_{\bm k} \equiv (a_{1,\bm{k}},\dots,a_{M,\bm{k}})^T$ is a column vector of bosonic annihilation operators, $a_{\alpha,\bm{k}}$ annihilates a boson with quasimomentum $\bm{k}$ in the $\alpha$ sublattice; % the Fourier-transformed of their position-space counterparts, $a_{\bm{x},\alpha} = \sum_{\bm{k}} a_{\bm{k},\alpha} e^{i\bm{k}\bm{x}}/\sqrt{N}$
$\bm{r}_n\in \mathbb{Z}_L^2$ and $\alpha_n\in\mathbb{Z}_M$ denote the unit cell and sublattice to which the $n$th emitter is coupled, respectively. 
% See Fig.~\ref{fig:schematic} for a schematic of the kind of systems under consideration.

\begin{figure}
    \centering
    \includegraphics{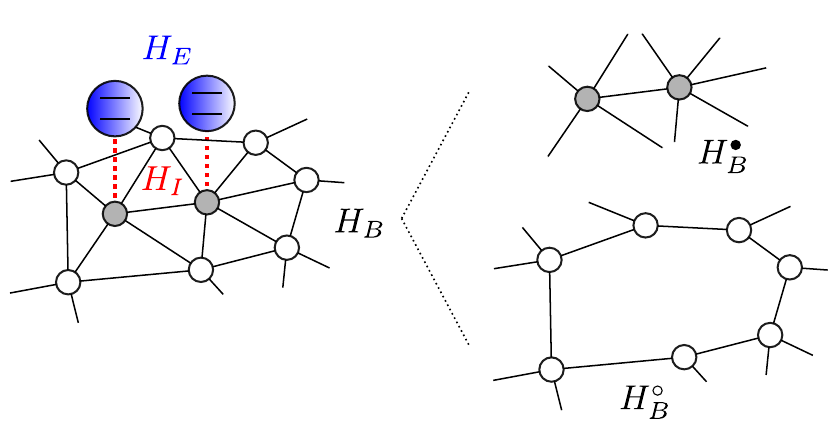}
    \caption{Schematic of the system under consideration. The bath can be split in two parts: a bath with vacancies, described by the Hamiltonian $H_B^\circ$, and the sites to which the emitters are coupled, described by $H_B^\bullet$.}
    \label{fig:schematic}
\end{figure}

For the bath, we will consider two paradigmatic examples of two-dimensional Chern insulators:% \cite{Asboth2016book}:

\emph{1. Harper-Hofstadter model} \cite{Hofstadter1976}. This model corresponds to a square lattice with nearest-neighbor hoppings of strength $J$, threaded by a magnetic flux. For a magnetic flux per plaquette $\phi = 2\pi p/q$ ($p,q\in \mathbb{N}$, coprime), the momentum-resolved bath Hamiltonian $\mathcal{H}_B(\bm{k})$, $\bm{k} =(k_x,k_y)$, can be represented by a $q\times q$ matrix, with elements
\begin{multline}
    \frac{1}{J}\left[\mathcal{H}_B(\bm{k})\right]_{mn} = -2\cos(k_x + n\phi)\delta_{m,n} - \delta_{m,n+1} \\ - \delta_{m,n-1} -e^{-ik_y}\delta_{n, m + q - 1}-e^{ik_y}\delta_{m, n + q - 1} \,.
\end{multline}

\emph{2. Haldane honeycomb model} \cite{Haldane1988}. 
This model corresponds to a honeycomb lattice with nearest-neighbor hoppings of strength $t_1$, and complex next-to-nearest-neighbor hoppings of strength $t_2e^{\pm i\phi}$. 
The complex phases are chosen such that there is no net magnetic flux threading the lattice, yet the model is not time-reversal invariant. 
In addition, there is a different on-site potential $\pm M$ for each sublattice. 
The momentum-resolved bath Hamiltonian can be expressed in terms of the Pauli matrices $\{\sigma^\nu\}_{\nu=x,y,z}$, and the identity matrix $\sigma^0$, as $\mathcal{H}_B(\bm{k}) = h_0(\bm{k}) \sigma^0 + \sum_\nu h_\nu(\bm{k})\sigma^\nu$, with
\begin{equation}
    \begin{split}
      h_0(\bm{k}) & = -2t_2\shortcos(\phi)\left[\shortcos(k_x) + \shortcos(k_y) + \shortcos(k_y - k_x)\right]\,,\\
      h_x(\bm{k}) & = -t_1\left[1 + \shortcos(k_x) + \shortcos(k_y)\right]\,,\\
      h_y(\bm{k}) & = -t_1\left[\shortsin(k_x) + \shortsin(k_y)\right]\,,\\
      h_z(\bm{k}) & = M + 2t_2\shortsin(\phi)\left[\shortsin(k_x) - \shortsin(k_y) + \shortsin(k_y - k_x)\right]\,,
    \end{split}
\end{equation}
where we have used the abbreviations $\shortsin(x)\equiv\sin(x)$ and $\shortcos(x)\equiv\cos(x)$.

\section{Vacancy-like bound states \label{sec:vbs}}

Since many of the effects discussed in this paper are due to the presence of vacany-like bound states (VBS) \cite{Leonforte2021}, in this section we briefly review their properties and translate them into properties of the emitter self-energy (also known as the level-shift operator \cite{Cohen1992atom}). 
Furthermore, we show their relevance in understanding the whole system's spectrum in the large-$g$ limit.

To define them, it is useful to split the bath Hamiltonian in two parts, $H_B = H_B^\circ + H_B^\bullet$, such that $H_B^\bullet$ contains only the terms involving the sites to which the quantum emitters are coupled, while $H_B^\circ$ contains the rest of the terms, see Fig.~\ref{fig:schematic} for a schematic representation.

\begin{theorem}[Existence of VBS, Ref.~\cite{Leonforte2021}]
\label{theo:vacancy-modes}
if $\ket{\Psi_\mathrm{VM}}$ is a single-particle eigenstate of $H^\circ_B$ with energy $E_\mathrm{VM}$, the state $\ket{\Psi_\mathrm{VBS}} = \sum_n c_{e,n} \sigma^+_n\ket{\mathrm{vac}} + \alpha \ket{\Psi_\mathrm{VM}}$ is an eigenstate of the whole Hamiltonian $H= H_E + H_B + H_I$ with the same energy $E_{\rm BS} = E_{\rm VM}$, provided $\Delta = E_\mathrm{VM}$.
\end{theorem}

We can identify vacancy modes with those energies $E_{\rm VM}$ for which the self-energy of the emitter subsystem $\Sigma$, is singular, $\det \Sigma(E_{\rm VM}) = 0$. 
Note that for a system in which $N$ emitters are coupled locally, each to a single site of a $D$-dimensional photonic lattice, the self-energy is a $N\times N$ matrix, whose elements are given by (see Appendix \ref{app:boundstates}) 
\begin{equation}
    \Sigma_{mn}(z) = g^2\left[\Phi(z,\bm{r}_m - \bm{r}_n)\right]_{\alpha_m,\alpha_n} \,, \label{eq:selfematrix}
\end{equation}
with $\Phi$ an $M\times M$ matrix given by 
\begin{equation}
    \Phi(z,\bm{r}) = \int_\mathrm{BZ} \frac{d^D\bm{k}}{(2\pi)^D}\, e^{i\bm{k}\cdot\bm{r}} \left[z - \mathcal{H}_B(\bm{k})\right]^{-1} \,, \label{eq:phi}
\end{equation}
where the integral is over the first Brillouin zone.
In particular, for a single emitter, the self-energy is a complex function, and the energies of the vacancy modes are its zeros on the real line, $\Sigma(E_{\rm VM}) = 0$. 
Theorem \ref{theo:vacancy-modes} follows from the fact that the bound-state energies satisfy the ``pole equation'' \cite{Leonforte2021nano}
\begin{equation}
    \det [E_{\rm BS} - \Delta - \Sigma(E_{\rm BS})] = 0 \,.
\end{equation}
Furthermore, one can show that the emitter weights of the bound states satisfy (Appendix \ref{app:boundstates}) 
\begin{equation}
    \bm{c}_e^\dag \left[1 - \Sigma'(\EBS)\right]\bm{c}_e = 1 \,,
\end{equation}
where $\Sigma'(z) \equiv \partial_z \Sigma(z)$, and $\bm{c}_e \equiv (c_{e,1},\dots,c_{e,N})^T$.

Theorem \ref{theo:vacancy-modes} holds regardless the value of the light-matter coupling constant $g$. 
As a consequence:
\begin{enumerate}[label=(\roman*)]
    \item The shape, in particular, the decay length, or the inverse participation ration (IPR), of the photonic component of the bound state, does not change as a function of $g$---it is fixed by $\ket{\Psi_\mathrm{VM}}$.
    \item The populations of the emitter excited states $\abs{c_{e,n}}^2$, decay monotonically like $\sim g^{-2}$ as $g\to\infty$.
\end{enumerate}
% These two features are unique to this kind of bound states. 
% For other kinds of bound states, $E_\mathrm{BS}$ is in general a function of $g$, which, in turn, makes $\xi$ to also depend on $g$, and $\abs{c_{e,n}}^2$ to have a different dependence on $g$.
% In Fig.~\ref{fig:singleQEbs} we show examples of these different behaviors for a single emitter coupled to the Harper-Hofstadter lattice.

\begin{figure}[h]
    \centering
    \includegraphics{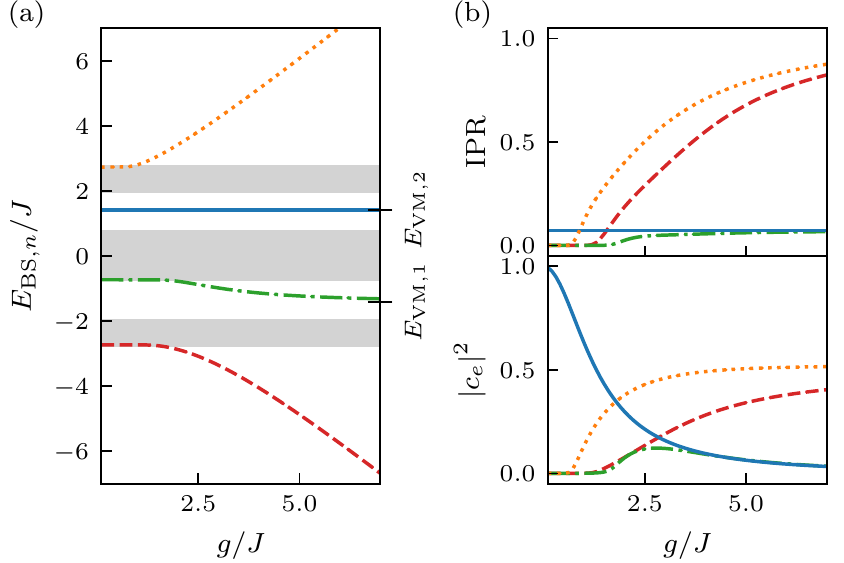}
    \caption{(a) Bound-state energies for a single emitter coupled to a single site of the Harper-Hofstadter model, with $\phi=2\pi/3$, as a function of the light-matter coupling strength. 
    Grey areas mark the range of the bath's energy bands, while the ticks on the right vertical axis mark the energies of the vacancy modes; $\Delta = E_{\rm VM, 2}$ (blue, solid line). 
    (b) Inverse participation ratio (IPR), and emitter's excited state occupation of the same bound states. 
    The IPR has been computed as $\mathrm{IPR}:=\sum_j \abs{\psi_j}^4$, where the sum runs over all bath sites, and $\psi_j$ denotes the probability amplitude to find the photon in the $j$th bath site, normalized such that $\sum_j \abs{\psi_j}^2 = 1$; $1/(L_x L_y) \leq \mathrm{IPR}\leq 1$. 
    The lowest value is attained for a homogeneous state (constant probability), while the maximum is attained for a fully localized state (at a single site). 
    A lattice of $(L_x, L_y) = (3 \cdot 2^4, 2^4)$ cells along each dimension, and periodic boundary conditions, has been used in this exact-diagonalization calculation.}
    \label{fig:singleQEbs}
\end{figure}

It should be noted that even if the condition $\Delta = E_\mathrm{VM}$ is not met, the presence of vacancy modes is relevant in the strongly-interacting regime $g\gg \Delta, \norm{H_B}$.
As shown in Fig.~\ref{fig:singleQEbs}(a), the bound-state energies may show two qualitatively different behaviors as the light-matter coupling increases: either they diverge or they converge to the energy of a vacancy mode. 
The former bound states, which we shall call \emph{strongly-hybridized polaritons}, are, in this regime, approximate eigenstates of the interaction Hamiltonian $H_I$. 
For our particular Hamiltonian, the eigenstates of $H_I$ are the symmetric and antisymmetric superpositions $\left(\sigma^+_n \pm a_{\bm{r}_n,\alpha_n}^\dag\right)\ket{\mathrm{vac}}/\sqrt{2}$. 
The latter kind is, in the same regime, approximately equal to the vacancy modes.
This explains the observed behavior of the emitter occupation $\abs{c_e}^2$, and the localization of the photonic component of the bound states, shown in Fig.~\ref{fig:singleQEbs}(b). 
Namely, for vacancy-like bound states, or bound states that tend to a vacancy-mode, $\abs{c_e}^2$ and the IPR show the behavior described in points (i) and (ii), whereas for strongly-hybridiced polaritons, $\abs{c_e}^2\to 1/2$ and ${\rm IPR} \to 1$ as $g\to\infty$.

\section{Few emitters \label{sec:few_emitters}}

% As we show in this section, for systems with a few emitters only it is difficult to find a qualitative feature of the bound states depending on the bath's topology.
Although the connection to the underlying bath topology in systems with few emitters is less straightforward than in the case of extended emitter arrangements, it is nevertheless interesting to look at the effects induced by the breaking of time-reversal symmetry and point out similarities and differences with other 2D baths.
As we show in this section, for the Harper-Hofstadter model, the major difference as compared with the square lattice model is the non-monotonicity of the bound states that appear in the gaps opened in the spectrum by the presence of a magnetic flux. 
For the Haldane model, similarly to the honeycomb lattice, a single emitter can seed a quasibound state that gives rise to true long-range interactions, which also break time-reversal symmetry at short distances. 

To show all this, it is instrumental to compute the emitter self-energy.
% As already hinted in the previous section, many of the properties of the emitters are determined by their self-energy. 
Knowing $\Sigma_{mn}(\Delta)$ for arbitrary distances $\bm{r}_{mn} \equiv \bm{r}_m - \bm{r}_n$ between the emitters and values of $\Delta$ outside the band ranges is %essentially 
equivalent to knowing the shape of the single-emitter bound states (see Appendix \ref{app:boundstates}), and thus also provides information about the effective coherent interactions mediated by the bath.
Unfortunately, obtaining analytical expressions for the self-energy is rather difficult for both the Harper-Hofstadter and Haldane models. 
A workaround is to integrate numerically its expression in terms of the momentum-resolved bath Hamiltonian, a task that can be facilitated by using as a starting point the exact expressions for the partial integrals shown in Appendix \ref{app:1D}.

\subsection{Harper-Hofstadter model}

% To connect with previous literature, and to point out similarities and differences with other 2D baths, in this section we briefly discuss the properties of the bound states and dynamics of few-emitter systems. 
The Harper-Hofstadter model has been partially covered in Ref.~\cite{Bernardis2021}, where they consider quantum emitters coupled to a square lattice threaded by a relatively small magnetic flux, one such that the magnetic length $l_B = l_0/\sqrt{\phi}$ is larger than the lattice constant $l_0$, but smaller than the lattice size, $1 < l_B/l_0 < \min(L_x,L_y)$.  
In this limit the bath spectrum splits into many sub-bands with a vanishingly small width (Landau levels); a single emitter resonant with one of the Landau levels does not decay, but its excited-state population oscillates as the photon hops back and forth between the emitter and the Landau orbital located around it.

\begin{figure}
    \centering
    \includegraphics{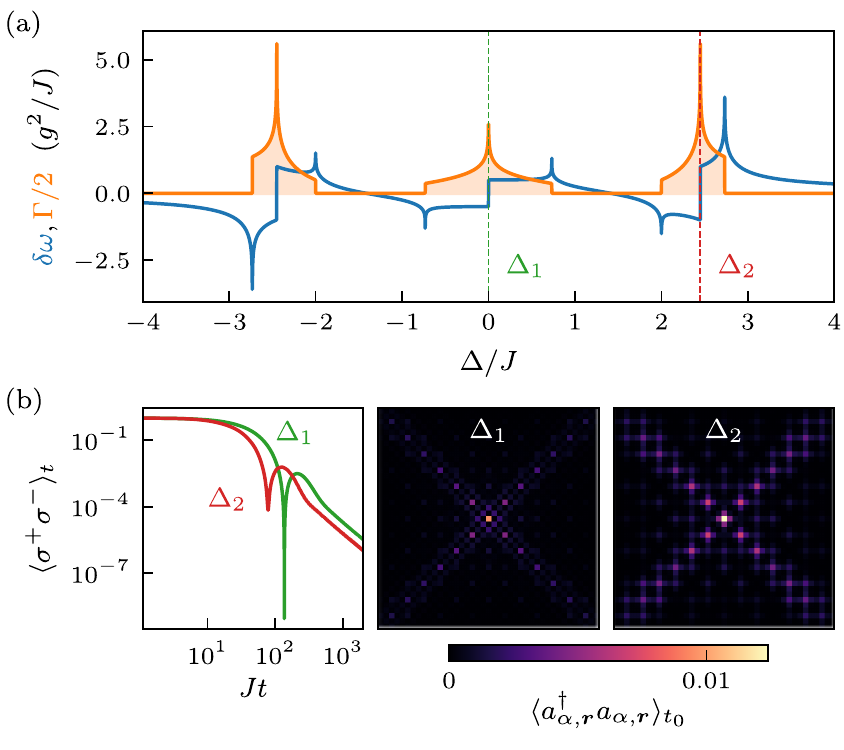}
    \caption{Self-energy (a), $\Sigma(\Delta + i0^+) = \delta\omega - i\Gamma/2$, and emission dynamics (b), for a single emitter coupled to the Harper-Hofstadter model with $\phi=2\pi/3$. 
    In (b) we show the emitter excited state occupation as a function of time for two different values of the emitter transition frequency $\Delta_j$ ($j=1,2$), tuned to different van-Hove singularities of the bath's density of states.
    Next to it we show respective snapshots of the bath occupation at time $t_0 = 50 J^{-1}$. 
    The coupling strength is in both cases $g=0.1J$.}
    \label{fig:harper-hofstadter_algebraic_decay}
\end{figure}

Here, instead, we consider the limit of large magnetic flux $l_B/l_0 < 1$, in which the bath spectrum consists only of a few sub-bands with a finite width. 
In this case, the emission dynamics is very similar to that found for the square lattice \cite{GonzalezTudela2017PRL,GonzalezTudela2017PRA}, provided the coupling strength is sufficiently small compared to the width of the sub-bands. 
The density of states for each sub-band has a similar shape, with a van-Hove singularity giving rise to the same long-term algebraic decay and emission profile peaked along the main diagonals of the lattice (see Fig.~\ref{fig:harper-hofstadter_algebraic_decay}).
Regarding the bound states, the main difference with respect to the square lattice is that they do not decay monotonically away from the emitter. 
This has important consequences for extended emitter arrays that we will review in the next section.

\subsection{Haldane model}

Haldane's model has been less explored in the quantum-emitter literature.
Here, we focus on the regime where both $\abs{M},\abs{t_2}\ll\abs{t_1}$, in which the Haldane model can be regarded as a small perturbation to the honeycomb lattice \cite{GonzalezTudela2018}.
We shall call this regime the ``honeycomb limit''.
In the following, we derive approximate expressions for the self-energy, valid in the honeycomb limit for values of $\Delta$ outside the band ranges, and analyze their implications for the effective bath-induced interactions between the emitters.

Particularizing Eqs.~\eqref{eq:selfematrix} and \eqref{eq:phi} to the case of a 2D, two-band bath, we can express it in integral form as
\begin{equation}
    \Sigma_{mn}(z) = g^2 \int_{\rm BZ} \frac{d^2\bm{k}}{(2\pi)^2}  \frac{e^{i\bm k \cdot \bm r_{mn}} f_{\alpha_m\alpha_n}(z,\bm k)}{\left[z - h_0(\bm k)\right]^2 - h^2(\bm k)} \,. \label{eq:generalselfe}
\end{equation}
Here, $h^2(\bm k)\equiv h_x^2(\bm k) + h_y^2(\bm k) + h_z^2(\bm k)$, 
% $\bm r_{mn}\equiv \bm r_m - \bm r_n$ denotes the signed distance between the $m$th and the $n$th emitter, 
and the numerator can take one of the following forms:
\begin{align}
    f_{AA}(z,\bm k) & =  z - h_0(\bm k) + h_z(\bm k)\,, \\
    f_{BB}(z,\bm k) & =  z - h_0(\bm k) - h_z(\bm k)\,, \\
    f_{AB}(z,\bm k) & =  h_x(\bm k) - i h_y(\bm k)\,, \\
    f_{BA}(z,\bm k) & =  h_x(\bm k) + i h_y(\bm k)\,,
\end{align}
depending on the sublattice to which the emitters are coupled.
The largest contribution to the integral appearing in  Eq.~\eqref{eq:generalselfe} comes from the bath modes whose frequencies are closest to that of the emitters. 
It is possible to obtain an analytical approximation expanding the integrand around the quasimomenta of such modes, if the distance between the emitters is large $\abs{\bm r_{mn}}\gg 1$ (see Appendix \ref{app:selfenergyHaldane}).
When $\Delta$ lies above or below the bath's bands and close to the band edges 
%, the expression we obtain for the self energy is real, and its spatial dependence is determined by zeroth-order Bessel function of the second kind, 
we obtain 
\begin{equation}
    % \Sigma_{mn} \propto \K_0(r'_{mn}/\xi_0) \,.
    \Sigma_{mn} \propto \K_0\left(\frac{r'_{mn}}{\xi_0}\right) \,.
\end{equation}
Here, $r'_{mn}$ is the Euclidean norm of $\bm r'_{mn}$, which is simply $\bm r_{mn}$ expressed in a different coordinate basis, and $\K_0$ denotes the zeroth-order Bessel function of the second kind. 
If instead, $\Delta$ lies in the gap between the two bands, 
\begin{equation}
    \Sigma_{mn}(\Delta)\simeq I_+(\Delta) + I_-(\Delta) \,,
\end{equation}
where each contribution comes from a different K-point, $\pm \bm K \equiv \pm(1, -1)2\pi/3$. 
For two emitters coupled to the same sublattice, 
\begin{equation}
    % I^\pm_{mn}(\Delta) \propto e^{\pm i \bm K \cdot \bm r_{mn}} \K_0(r'_{mn}/\xi_\pm) \,,
    I^\pm_{mn}(\Delta) \propto e^{\pm i \bm K \cdot \bm r_{mn}} \K_0\left(\frac{r'_{mn}}{\xi_\pm}\right) \,,
\end{equation}
while for two emitters coupled to different sublattices, say $(\alpha_m, \alpha_n) = (A, B)$, 
\begin{equation}
    % I^\pm_{mn}(\Delta) \propto e^{\pm i (\bm K \cdot \bm r_{mn} - \theta_{mn} - \pi/2)} \K_1(r'_{mn}/\xi_\pm)/\xi_\pm \,.
    I^\pm_{mn}(\Delta) \propto \frac{e^{\pm i (\bm K \cdot \bm r_{mn} - \theta_{mn} - \pi/2)}}{\xi_\pm} \K_1\left(\frac{r'_{mn}}{\xi_\pm}\right)\,. \label{eq:long-range}
\end{equation}
Here, $\theta_{mn}$ denotes the angle between $\bm r'_{mn}$ and the horizontal axis, and $\K_1$ denotes the first-order modified Bessel function of the second kind.

\begin{figure}
    \centering
    \includegraphics{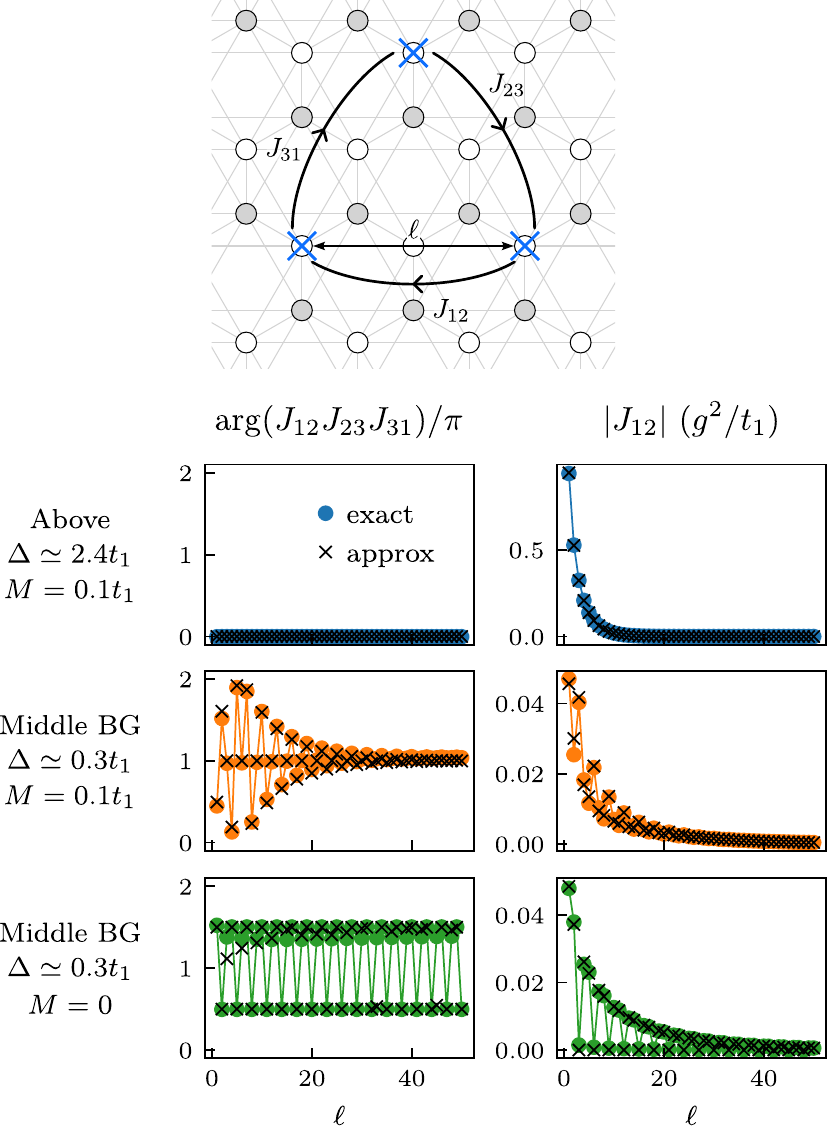}
    \caption{Strength of the effective interactions (right column) and total magnetic flux (left column) threading a triangular emitter plaquette, as shown in the schematics above, % located at positions $\bm r_1 = (0, 0)$, $\bm r_2 = (\ell, 0)$, $\bm r_3 = (0, \ell)$, 
    as a function of the distance $\ell$ between the emitters. % $\ell = r'_{nm}\sqrt{3}/2$, for $n\neq m$. 
    The dots have been obtained by numerical integration of Eq.~\eqref{eq:selfematrix}, while the crosses correspond to the approximate analytic formulas (detailed in Appendix \ref{app:selfenergyHaldane}). 
    The remaining parameters are $t_2 = 0.1 t_1$ and $\phi = 0.1$, the same in all the cases shown.}
    \label{fig:flux_triangle}
\end{figure}

In light of these expressions we conclude that for emitters tuned above and below the bands the effective interactions mediated by the bath are approximately real.
In the middle band gap, $I^+_{mn}(\Delta)$ and $I^-_{mn}(\Delta)$ may have different magnitude, resulting in complex effective interactions. 
This is a consequence of breaking the symmetry between the two different K-points, which, by contrast, is always present in the honeycomb lattice.
Individual phases do not determine by themselves whether the polariton dynamics is time-reversal invariant or not.
Instead, a gauge-invariant quantity quantifying this is the total phase accumulated when hopping from one emitter to another in a closed path, or, in other words, the flux threading the path.
% One may wonder to what extent the emitter dynamics induced by the bath (with broken time-reversal symmetry) breaks itself time-reversal symmetry.
In Fig.~\ref{fig:flux_triangle}, we present this calculation for three emitters forming a triangle, for different values of the detuning: above the bath's bands, and in the middle band gap.
% To quantify this, we compute the flux threading a plaquette of emitters, shown in Fig.~\ref{fig:flux_triangle}.
As it turns out, the flux tends to zero as the distance between the emitters increases. 
Only when the symmetry between the two K-points is restored ($M = 0$), the interaction ranges are equal $\xi_+ = \xi_-$, and the flux does not decay, but it oscillates between $\pm \pi/2$, which corresponds to perfectly chiral dynamics \cite{Bernardis2021}. 

% In general, $I^+_{mn}(\Delta)$ and $I^-_{mn}(\Delta)$ may have different magnitude, resulting in complex effective interactions.

% One might expect a time-reversal-symmetry-broken bath to always produce an emitter dynamics that also breaks time-reversal symmetry. 
% However, this is not always the case.
% A gauge-invariant quantity quantifying this is the effective flux threading a plaquette of emitters.
% For example, we can consider three emitters coupled to the same sublattice, forming an equilateral triangle.
% Using the approximate formulas described previously, we realize that the total flux vanishes as we increase the distance between the emitters, except in some specific situations. 
% For detunings above and below the bands, this is clear from the fact that the self-energy is real. 
% For detunings in the gap, the two interaction decay lengths $\xi_\pm$ are different in general, such that at long distances only one of the contributions, $I^+_{mn}(\Delta)$ or $I^-_{mn}(\Delta)$, dominates, and the sum of the phases picked when traversing the loop in a clockwise or anticlockwise manner vanishes. 
% When $M = 0$, $\xi_+ = \xi_-$, and the resulting flux turns out to oscillate between $\pm \pi/2$, as shown in Fig.~\ref{fig:flux_triangle}, which corresponds to perfectly chiral dynamics \cite{Bernardis2021}.

% is essentially a Fourier transform of a function $F$. 
% If the distance between the emitters is large $\abs{\bm r_{mn}}\gg 1$, the integrand is highly oscillatory, such that only the sharpest features of $F$ give a substantial contribution. 
% These extrema occur precisely around the quasimomenta of the modes whose frequencies are closest to that of the emitters.
% Thus, in this regime we can approximate the integral by replacing the numerator and denominator in $F$ by some truncated series expansion around these extremal points (see Appendix \ref{app:selfenergyHaldane}).

% When both $\abs{M},\abs{t_2}\ll \abs{t_1}$, the additional terms of the Haldane model can be though of as a small perturbation to a regular honeycomb lattice.
% In this regime, which we shall call the ``honeycomb limit'', the maximum (minimum) of the upper (lower) band occurs at the $\Gamma$-point, $\bm \Gamma \equiv (0, 0)$, whereas the minimum (maximum) of the upper (lower) band occur at either one of the K-points, $\pm \bm K \equiv \pm (1, -1) 2\pi/3$.
% Thus, when $\Delta$ lies above or below the bands, it is sufficient to expand $F$ around the $\Gamma$-point.
% To leading order, we obtain a real value for the self energy, $\Sigma_{mn} \propto \K_0(r'_{mn}/\xi_0)$. 
% Here, $\K_0$ denotes the zeroth-order modified Bessel function of the second kind, and $r'_{mn}\equiv \norm{\bm r'_{mn}}$, where $\bm r'_{mn}$ is $\bm r_{mn}$ expressed in a different coordinate basis. 
% the euclidean distance between the emitters.

\begin{figure}
    \centering
    \includegraphics{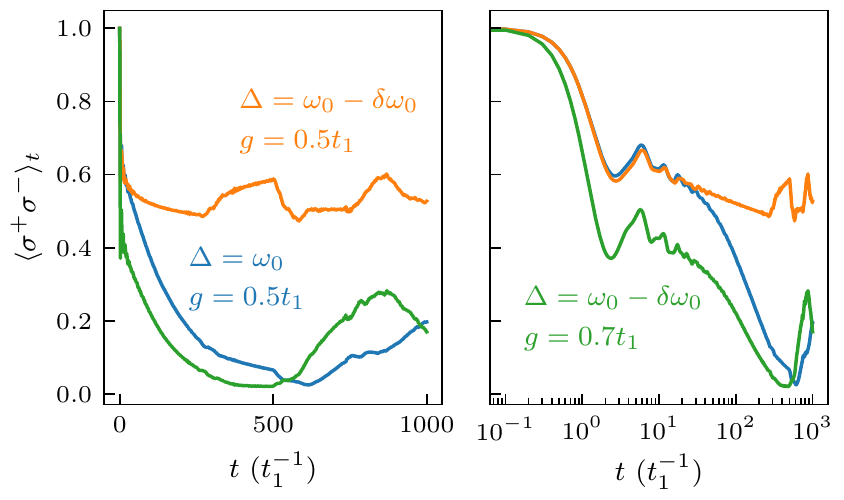}
    \caption{Emitter excited state occupation as a function of time for three different configurations, showcasing the sensitivity of the dynamics to the emitter detuning and light-matter coupling strength. 
    The bath parameters are the same in all cases and correspond to a critical bath with a Dirac point at $-\bm K$: $t_2 = 0.1t_1$, $\phi = 0.5$, and $M = 3\sqrt{3}t_2\sin(\phi)$. 
    The Dirac-point frequency is denoted by $\omega_0$, while $\delta\omega_0$ denotes the Lamb shift at $\omega_0$ for $g = 0.5 t_1$. 
    The simulation has been carried out in a lattice with $L_x = L_y = 2^8$ unit cells along each dimension and periodic boundary conditions.
    The right-hand side plot shows the same data in logarithmic scale.}
    \label{fig:cone_decay}
\end{figure}

For a critical bath, i.e., when the gap closes at either $\pm\bm K$, the respective interaction range diverges, ${\xi_\pm \to \infty}$. 
From Eq.~\eqref{eq:long-range}, using the asymptotic relation $\K_1(x)\sim 1/x$ for $x\to 0$, we realize that the effective interaction between two emitters coupled to different sublattices decays as $\sim 1/r'_{mn}$ when they are spectrally tuned to the vanishing gap, so it is truly long-range.
The same thing happens in the honeycomb lattice, which presents two Dirac cones at both $\pm\bm K$, and has therefore a singular gap.
When the emitter transition is tuned to the frequency of the singular gap a quasibound state appears, which scales as $\sim 1/r$, where $r$ is the distance to the emitter, giving rise to long-range effective interactions \cite{GonzalezTudela2018}.
This quasibound state is, in the case of the honeycomb lattice, actually a vacancy-like quasibound state, meaning that its energy and photonic component remain the same, regardless the value of $g$.
In the Haldane model, the origin of the long-range interactions can be attributed also to a quasibound state, although in this case it is not of the vacancy-like type. 
In the Haldane model the gap does not close simultaneously at both K-points, and as a consequence, the self-energy for a single emitter does not vanish at the frequency of the single Dirac point; while the density of states indeed vanishes, the Lamb shift is different from zero, so there is no longer a vacancy mode at that frequency.
Instead, there will be a quasibound state whenever the detuning $\Delta$ compensates precisely for the Lamb shift.
In such a case, an emitter initially excited will not decay completely, as shown in Fig.~\ref{fig:cone_decay}, even though its frequency is resonant with some guided modes in the bath. 
This quasibound state is fragile, in the sense that it only exists for particular values of $(\Delta, g)$.
From the approximate expressions for the self-energy, we can see that at long distances it also scales like $\sim 1/r$, and has support mostly on the opposite sublattice to which the emitter is coupled.

\section{1D emitter arrangements \label{sec:line_defects}}

The main characteristic of 2D Chern insulators is the presence of chiral edge states in lattices with open boundary conditions (OBC) \cite{Hatsugai1993}. 
Equivalently, we can consider lattices with vacancies forming a 1D (line) defect. 
Since the energies of the chiral edge states span the whole range of a bandgap, we are guaranteed to find vacancy-like bound states for arrangements of emitters forming a line defect, provided the emitter transition frequency lies in one of the topological bandgaps. 
Furthermore, since the number of edge states is linked to the Chern numbers of the bands above and below the gap, the number of vacancy-like bound states that can be found is also determined by the Chern numbers in general. 
In the following, we demonstrate this effect in the Harper-Hofstadter model and the Haldane model. 

\subsection{Harper-Hofstadter model}

\begin{figure*}
    \centering
    \includegraphics{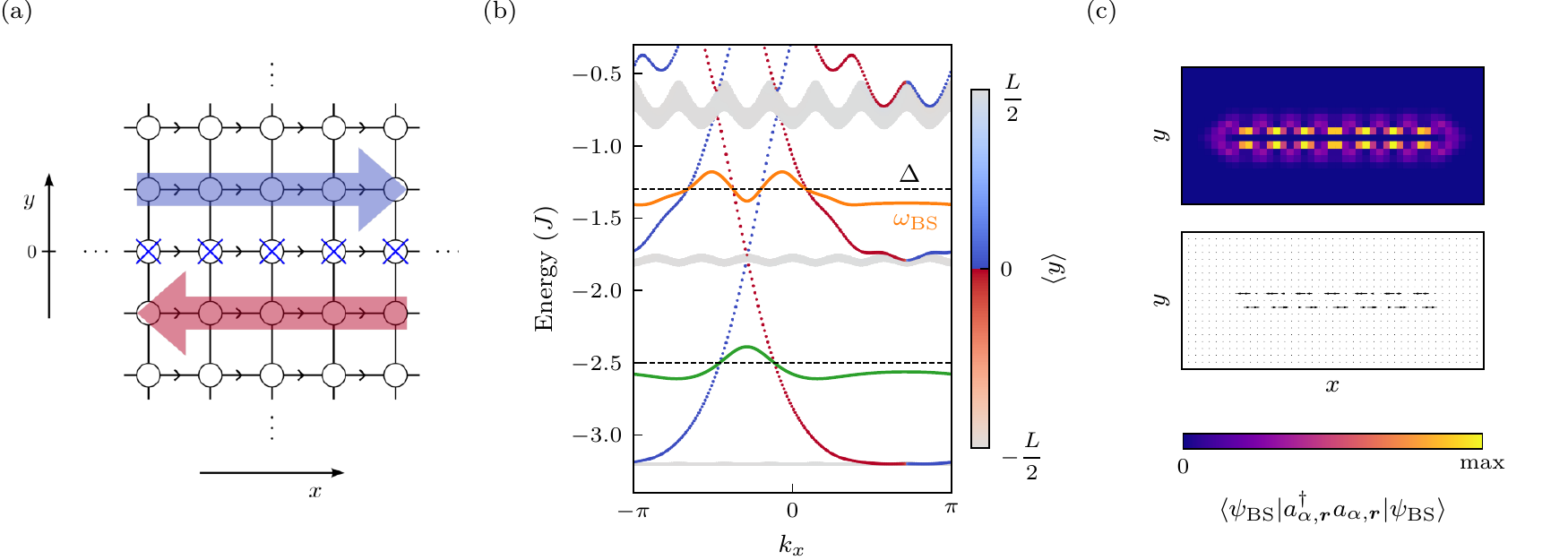}
    \caption{(a) Schematic representation of an emitter line defect in the Harper-Hofstadter model. 
    The dots represent the bath sites, while lines connecting them represent hopping processes (marked with an arrow if they are complex).
    Blue crosses mark the sites to which the quantum emitters are coupled. 
    The big, red and blue arrows represent chiral edge states that are present in the model with vacancies.
    (b) Spectrum of $H_B^\circ$ for the same system, with $\phi = 2\pi/7$. 
    The spectrum is color-coded according to the localization of the eigenstates with respect to the line defect. 
    On it, we show the polariton bands (orange and green lines) obtained for two different values of $\Delta$ (dashed horizontal lines). Vacancy-like bound states correspond to the crossings between the edge-state energies and $\Delta$.
    The light-matter coupling constant is $g=0.5J$.
    (c) Photonic component of the bound state with energy closest to $\Delta$ for a finite emitter line defect. 
    The parameters are the same as in panel (b), with $\Delta = -1.3J$. 
    In addition to the probability of finding the photon in a given bath site (color plot), we also show the probability current \cite{Boykin2010}.}
    \label{fig:harper-hofstadter_1D}
\end{figure*}

Let us consider first a set of quantum emitters coupled to consecutive sites of the Harper-Hofstadter model along the $x$-direction, i.e., their coordinates in the basis of the lattice vectors are $\bm{r}_n\equiv(x_n, y_n) = (n, 0)$, see Fig.~\ref{fig:harper-hofstadter_1D}(a) for a schematic representation. 
The presence of emitters breaks the bath's  translational symmetry in the direction perpendicular to the line defect, but we can still Fourier-transform in the parallel direction, obtaining a collection of 1D single-emitter systems parametrized by the quasimomentum $k_x$, $H=\sum_{k_x} H^\mathrm{1D}(k_x)$, with 
\begin{equation}
    H^\mathrm{1D}(k_x) = \Delta \sigma^+_{k_x}\sigma^-_{k_x} + g(\sigma^+_{k_x} a_{k_x,0} + \mathrm{H.c.}) + H_B^\mathrm{1D}(k_x) \,. \label{eq:H1D}
\end{equation}
In this case, $H_B^\mathrm{1D}$ corresponds to a chain with nearest-neighbor hopping of amplitude $J$, and periodic on-site potentials $\mu_n = -2J\cos(k_x + n\phi)$ (see Appendix \ref{app:1DHarperHofstadter}).
Due to their proximity, the bound states of neighboring emitters will hybridize forming a band. 
Its dispersion relation $\omega_\mathrm{BS}(k_x)$, can be obtained solving the corresponding pole equation for each 1D system,
\begin{equation}
    \omega_\mathrm{BS} - \Delta - \Sigma(\omega_\mathrm{BS}; k_x) = 0\,. \label{eq:pole1d}
\end{equation} 
Here, $\Sigma(z; k_x)$ is the single-emitter self-energy corresponding to the 1D system described by Eq.~\eqref{eq:H1D}, which can be computed analytically (see Appendix \ref{app:1DHarperHofstadter}).

According to Theorem \ref{theo:vacancy-modes}, whenever $\Delta$ matches the energy of an edge state (for a given $k_x$), there will be a vacancy-like bound state. 
Thus, the number of vacancy-like bound states for any $\Delta$ inside the $n$th gap is, in general, equal to the number of edge modes crossing that gap, see Fig.~\ref{fig:harper-hofstadter_1D}(b). 
In other words, for $\max \omega_{n-1} < \Delta < \min \omega_n$, the number of zeros of $\Sigma(\Delta; k_x)$ as a function of $k_x$, $\zeta(\Delta)$, is lower bounded by $\zeta(\Delta)\geq 2(c_n - c_{n-1})$. 
Here $\omega_n$ and $\omega_{n-1}$ denote the dispersion relations of the bath bands above and below the $n$th gap, respectively, and $c_n$ and $c_{n-1}$ denote their corresponding Chern numbers. 
The factor 2 is due to the fact that we have to take into account the edge states on both sides of the line defect, which have opposite chirality. 
The only cases in which this bound is not satisfied is when $\Delta$ is tuned precisely to a crossing between edge modes. 

For an infinitely long emitter line defect, or for a system with periodic boundary conditions (PBC), the photonic component of these vacancy-like bound states is actually a chiral edge state of the bath. 
And therefore, it has support only in one of the two half-planes into which the line defect splits the bath. 
However, for sufficiently large but finite line defects, the true bound states are linear combinations of the (degenerate) bound states found for PBC, combined in such a way that they meet the OBC. 
For the Harper-Hofstadter model, degenerate vacancy-like bound states happen to have exactly the same decay length, but opposite chirality and directionality, so the vacancy-like bound state ends up decaying symmetrically away from the emitter line defect, see Fig.~\ref{fig:harper-hofstadter_1D}(c). 

\begin{figure}
    \centering
    \includegraphics{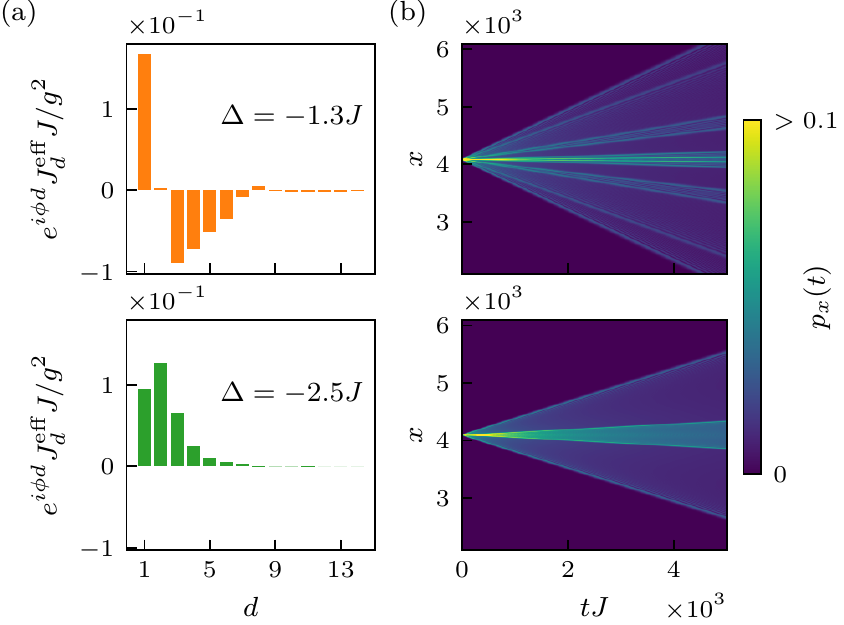}
    \caption{(a) Effective hopping amplitudes corresponding to the polariton bands shown in Fig.~\ref{fig:harper-hofstadter_1D}(b).
    (b) Ballistic diffusion of an initially localized polariton for the two different polariton bands.}
    \label{fig:hoppings_and_difussion}
\end{figure}

As we can appreciate in Fig.~\ref{fig:harper-hofstadter_1D}(b), not only the number of zeros of $\Sigma(\Delta,k_x)$ depends on the topology, but the overall ``bumpiness'' of the bound-state dispersion relation increases for increasing number of edge states. 
We can write an effective tight-binding model for bound states localized around the emitter positions as $H_\mathrm{eff} = \sum_{n,\, d} J^\mathrm{eff}_d \beta^\dag_{n + d} \beta_n$, where $\beta^\dag_n$ and $\beta_n$ create and annihilate a bound state (polariton) localized around the $n$th emitter, respectively. 
The hopping constants can be obtained from the dispersion relation as $J^\mathrm{eff}_d = \sum_k \omega_{\mathrm{BS}}(k) e^{ikd}/N$. 
Since $H^{\rm 1D}_B$ is symmetric with respect to the change $k_x - m\phi\to -k_x - m\phi$, where $m\in\{1,2,\dots, q\}$ labels the sublattice to which the emitters are coupled, the same kind of inversion symmetry is present in all the spectrum, which, in turn, implies that the effective hopping constants are real up to a phase, $\imag(e^{i m \phi d} J^{\rm eff}_d) = 0$, which can be absorbed in the definition of the polariton creation and annihilation operators.
The ``bumpiness'' of the dispersion relation translates into exotic hopping constants, which, although decaying with an overall exponential factor, may be larger for longer distances than for shorter distances, see Fig.~\ref{fig:hoppings_and_difussion}(a). 
This can be observed, e.g., by letting an initially localized polariton, $\ket{\psi(0)} = \beta^\dag_0\ket{\mathrm{vac}}$, evolve for a short period of time. 
The resulting wave function, $\ket{\psi(\delta t)}\simeq (1 - i \delta t H_\mathrm{eff}) \ket{\psi(0)}$, essentially maps the hopping amplitudes to occupations of the neighboring sites. 
The peculiar shape of the effective hoppings can be traced back to the particular shape of the single-emitter bound states \cite{Matsuki2021}, which in the continuum limit ($\phi\to 0$) have the nodes typical of Landau orbitals \cite{Bernardis2021}. 

The particular shape of $\omega_\mathrm{BS}$ also has visible consequences in the ballistic transport of a localized excitation. 
For a localized initial state, the wave function develops several wavefronts, which propagate linearly in time with a critical group velocity $v_c$, see Fig.~\ref{fig:hoppings_and_difussion}(b). 
This can be understood from the behavior of the probability $p_x(t) := \bra{\psi(t)} \beta^\dagger_x \beta_x \ket{\psi(t)}$ at large space-time scales. 
In the limit $x, t\to\infty$, keeping constant the ratio $x/t = \xi$, it is possible to approximate it using the stationary-phase approximation method \cite{Bender1999},
\begin{equation}
\begin{split}
    p_\xi(t) & = \abs*{\int_{\rm BZ} \frac{dk}{2\pi}\, e^{i[k\xi-\omega_{\rm BS}(k)]t}}^2 \\
    & \sim \abs*{\sum_{k_0} \frac{e^{i f(k_0) t + \sign\left(f''(k_0)\right) i \pi/4}}{\sqrt{2\pi t\abs{f''(k_0)}}} + O(t^{-1})}^2 \,.
\end{split}
\end{equation}
Here, the sum runs over all stationary points of the phase $f(k) = k\xi - \omega_{\rm BS}(k)$, $f'(k_0) = 0$, within the integration range. 
Clearly, the asymptotic expansion diverges at $\xi = v_c$, since there the second derivative of $f$ vanishes, $f''(k_0) = \omega_{\rm BS}''(k_0) = 0$.

These features of 1D arrangements of emitters only appear in the Markovian regime.
For $\norm{H_I} \gg \norm{H_B}, \norm{H_S}$, the effective hopping amplitudes decay monotonically as a function of distance, and initially-localized excitations will diffuse with just a single wavefront (see Appendix \ref{app:strong_light-matter}).
This can be understood noting that the single-emitter bound states appearing below (above) the lowest (highest) bath band decay monotonically as we move away from the emitter, and the strongly-hybridized polaritons have energies in those ranges.

\subsection{Haldane model}

\begin{figure*}
    \centering
    \includegraphics{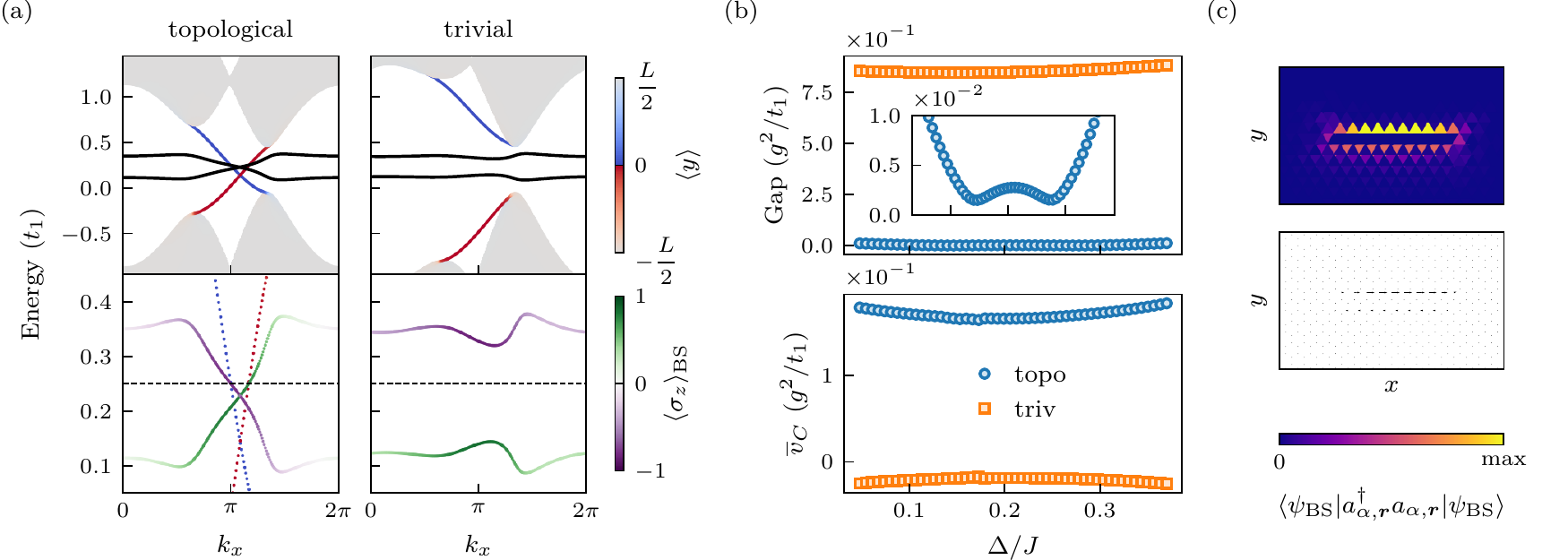}
    \caption{(a) Energy spectrum of $H_B^\circ$, and polariton bands $\omega_{\mathrm{BS},\pm}$ (black lines) for a set of emitters coupled to the Haldane model forming a line defect. 
    The parameters of the bath are $t_2 = 0.1t_1$, $\phi=0.8$ and $M\simeq 0.11t_1$ (topological) or $M\simeq 0.63t_1$ (trivial).
    The spectrum of $H_B^\circ$ is color-coded according to the localization of the eigenstates with respect to the line defect. 
    Below we show a zoom to the polariton bands, which are color-coded according to the weight of the eigenstates in each sublattice, $\sigma_z \equiv \sigma^+_A\sigma^-_A - \sigma^+_B\sigma^-_B$. 
    The dashed horizontal line marks the value of $\Delta\simeq 0.25 t_1$; and $g=0.5t_1$.
    (b) $\mathrm{Gap}:=\min_{k_x}\abs{\omega_{\mathrm{BS},+}(k_x) - \omega_{\mathrm{BS},-}(k_x)}$, and average chiral group velocity, Eq.~\eqref{eq:acgv}, for the same values of the bath parameters and varying emitter detuning; $g=0.1t_1$.
    The inset in the upper plot is a zoom to the values in the topological case.
    (c) Bound state with energy closest to $\Delta$, for a finite emitter line defect in a topological bath with the same parameters as in panel (a).}
    \label{fig:haldane_1D}
\end{figure*}

Now, we turn our attention to the Haldane model. 
The setup we analyze is essentially the same as for the Harper-Hofstadter model, shown in Fig.~\ref{fig:harper-hofstadter_1D}(a), albeit in this case we consider emitters coupled to both sites of each unit cell along a given direction, in order to make $H_B^\circ$ correspond to a bath with open boundary conditions. 
Thus, $H^\mathrm{1D}(k_x)$ corresponds in this case to a ladder with two emitters, each coupled to one of the sites in a given unit cell. 
As it happened for the Harper-Hofstadter model, the self-energy matrix can also be computed analytically in this case (see Appendix \ref{app:1DHaldane}). 
Since there are two emitters per unit cell of the line defect, we obtain two bands for the polariton modes, which are the solutions of
\begin{equation}
    \det\left[\omega_{\mathrm{BS},\pm} - \Delta - \Sigma(\omega_{\mathrm{BS},\pm};k_x)\right] = 0\,, \label{eq:pole1d2emitter}
\end{equation}
or equivalently, $\omega_{\mathrm{BS},\pm} - \Delta - \lambda_\pm(\omega_{\mathrm{BS},\pm}; k_x) = 0$, where $\lambda_\pm(\omega; k_x)$ denote the eigenvalues of the self-energy matrix $\Sigma(\omega; k_x)$,
\begin{equation}
    \lambda_\pm = \frac{1}{2}\left[\Sigma_{11} + \Sigma_{22} \pm \sqrt{(\Sigma_{11} - \Sigma_{22})^2 - 4\abs{\Sigma_{12}}^2}\right] \,. \label{eq:eigvals}
\end{equation}
Again, for any crossing of $\Delta$ with the edge-mode frequencies, we obtain a perfect vacancy-like bound state, see Fig.~\ref{fig:haldane_1D}(a). 
Note that degeneracies of the edge states translate into degeneracies of the vacancy-like bound states. 
Thus, for topologically non-trivial Haldane baths, there exist a value of $\Delta$---tuned to the crossing of the edge states---such that the resulting polariton bands are degenerate for some $k_x$. 
In general, we cannot expect degeneracies of the polariton bands to occur for a range of values of $\Delta$ and $k_x$, since, according to Eq.~\eqref{eq:eigvals}, they require that both $\Sigma_{11} = \Sigma_{22}$ and $\Sigma_{12} = 0$ simultaneously. 
In practice, however, we observe that the effective polariton bands remain almost gapless regardless the value of $\Delta$, as long as it lies in a topologically non-trivial gap, see Fig.~\ref{fig:haldane_1D}(b). 
By contrast, for the same emitter arrangement, the polariton bands are always gapped if the bath is in the trivial phase.
A difference with respect to the Harper-Hofstadter model is that now the bound states for finite line defects do not have to be symmetric with respect to the emitter line, as shown in Fig.~\ref{fig:haldane_1D}(c).

\begin{figure}
    \centering
    \includegraphics{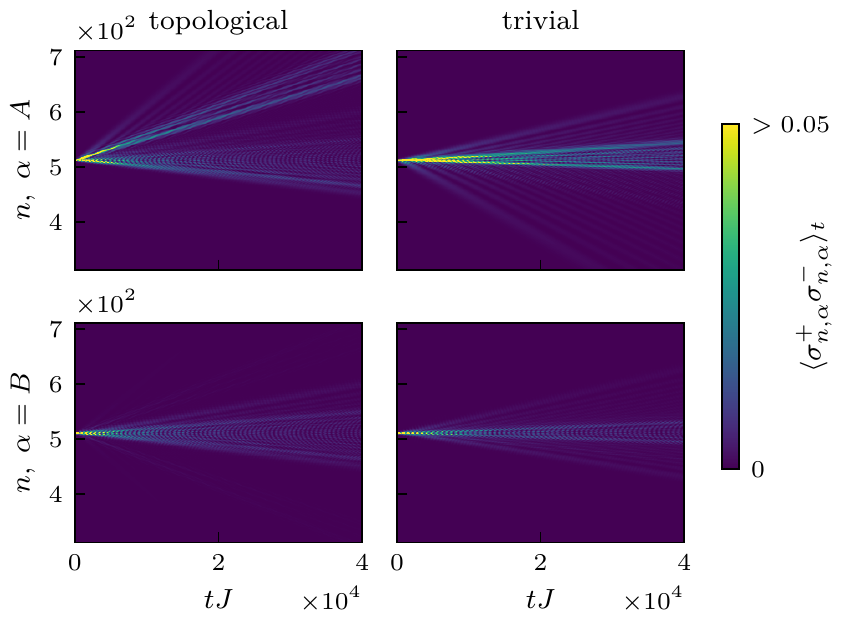}
    \caption{Emitter dynamics for an initial state consisting of a single excitation localized in sublattice $A$, $\ket{\psi(0)} = \sigma^+_{L/2, A} \ket{\rm vac}$. 
    The parameters of the system are $\Delta\simeq 0.25t_1$, $g=0.1t_1$, $t_2 = 0.1t_1$, $\phi=0.8$ and $M\simeq 0.11t_1$ (topological) or $M\simeq 0.63t_1$ (trivial).}
    \label{fig:sublattice_dynamics}
\end{figure}

Another difference with respect to the case studied in the Harper-Hofstadter model is the sublattice degree of freedom present in the emitter line defect, which allows for truly complex effective hoppings. 
One of the consequences of this, is the chiral dynamics that occur when an excitation is initially localized in one of the emitter sublattices, as demonstrated in Fig.~\ref{fig:sublattice_dynamics}.
We can understand this effect computing the average chiral group velocity $\overline{v}_C$ \cite{Huegel2014}, which quantifies the average group velocity of the polaritons in the system, weighted by their relative occupation in each emitter sublattice. 
Specifically,
\begin{equation}
    \overline{v}_C = \frac{1}{2}\sum_{\alpha = \pm} \int_{-\pi}^\pi \frac{dk}{2\pi}  v_\alpha(k)\mean{\sigma_z}_{k,\alpha} \,, 
    \label{eq:acgv}
\end{equation}
where
\begin{equation}
    \mean{\sigma_z}_{k,\alpha} \equiv \bra{\psi_{\mathrm{BS},\alpha}(k)} \sigma^+_{A,k}\sigma^-_{A,k} - \sigma^+_{B,k}\sigma^-_{B,k} \ket{\psi_{\mathrm{BS},\alpha}(k)}\,,
\end{equation}
and $v_\alpha(k) = \partial_k \omega_{\mathrm{BS},\alpha}(k)$. 
As can be seen in Fig.~\ref{fig:haldane_1D}(b), the value of $\overline{v}_C$ for line defects in topological baths is much larger than in the case of trivial baths.
This chiral dynamics is lost if the light-matter coupling constant is too large (see Appendix \ref{app:strong_light-matter}). 

\section{2D emitter arrangements \label{sec:emitter_superlattices}}

Going beyond 1D line defects, we can arrange the emitters periodically in the two dimensions of the bath forming what we may call an ``emitter superlattice''. 
Such systems have been considered in the past as a platform where one could generate strongly-correlated states of light and matter \cite{Greentree2006}, fractional quantum Hall states \cite{Hayward2012}, or more generally, use them as quantum simulators \cite{Hartmann2007,Hung2016,Manzoni2017,Bello2022}.
One might expect the properties of the resulting polariton models, in particular their topology, to depend strongly on the properties of the underlying bath. 
In this section we show how this is not the only aspect to take into account, but the geometry of the emitter arrangement and the coupling strength $g$ also play a crucial role.

First, we consider the case in which there is an emitter coupled to every site of every unit cell in some region of the bath. 
If $U_{\bm{k}}$ is a unitary that diagonalizes the bath Hamiltonian, $H_B = \sum_{\bm{k}} \tilde \Psi_{\bm{k}}^\dag \Omega(\bm{k}) \tilde \Psi_{\bm{k}}$, with
\begin{align}
    \Omega(\bm{k}) & \equiv U_{\bm{k}}\mathcal{H}_B(\bm{k})U^\dag_{\bm{k}} = \diag\left(\omega_1(\bm{k}),\dots,\omega_M(\bm{k})\right) \,, \\
    \tilde \Psi_{\bm{k}} & \equiv U_{\bm{k}} \Psi_{\bm{k}} = [\tilde a_{\bm{k},1},\dots,\tilde a_{\bm{k},M}]^T \,,
\end{align}
we can apply the same transformation to the emitter raising and lowering operators, defining $[\tilde \sigma^-_{\bm{k},1},\dots,\tilde \sigma^-_{\bm{k},M}]^T = U_{\bm{k}}[\sigma^-_{\bm{k},1},\dots,\sigma^-_{\bm{k},M}]^T$, such that the whole system Hamiltonian can be expressed as $H=\sum_{j=1}^M\sum_{\bm{k}} H_j(\bm{k})$, with
\begin{multline}
    H_j(\bm{k}) =  \Delta\tilde\sigma^+_{\bm{k},j}\tilde\sigma^-_{\bm{k},j} + \omega_j(\bm{k}) \tilde a^\dag_{\bm{k},j}\tilde a_{\bm{k},j} \\ + g(\tilde\sigma^+_{\bm{k},j}\tilde a_{\bm{k},j} + \mathrm{H.c.}) \,.
\end{multline}
Since $H_j(\bm{k})$ only involves a single mode and a single emitter raising/lowering operator, it can be easily diagonalized, yielding the following eigenmodes and eigenenergies:
\begin{gather}
    \lambda_{j,\pm}(\bm{k}) = \frac{\Delta + \omega_j(\bm{k})}{2} \pm \sqrt{\left(\frac{\Delta - \omega_j(\bm{k})}{2}\right)^2 + g^2}\,, \label{eq:eigenvalues}\\
    \beta^\dag_{\bm{k},j,+} = \cos\frac{\theta_{\bm{k}}}{2}  \tilde\sigma^+_{\bm{k},j} + \sin\frac{\theta_{\bm{k}}}{2}\tilde a^\dag_{\bm{k},j}\,,\\
    \beta^\dag_{\bm{k},j,-} = \sin\frac{\theta_{\bm{k}}}{2}  \tilde\sigma^+_{\bm{k},j} - \cos\frac{\theta_{\bm{k}}}{2}\tilde a^\dag_{\bm{k},j}\,,\\
    \cos\theta_{\bm{k}} = \frac{\Delta - \omega_j(\bm{k})}{\sqrt{[\Delta - \omega_j(\bm{k})]^2 + 4g^2}} \,.
\end{gather}
We can see that in the Markovian regime, $g\ll \abs{\Delta - \omega_j(\bm{k})}$, $\lambda_{j,+}\simeq \Delta$, $\lambda_{j,-}\simeq \omega_j(\bm{k})$, $\beta_{\bm{k},j,+}\simeq \tilde\sigma_{\bm{k},j}$ and $\beta_{\bm{k},j,-}\simeq \tilde a_{\bm{k},j}$, assuming $\Delta > \omega_j(\bm{k})$ (the same is true also for $\Delta < \omega_j(\bm{k})$, interchanging the labels $+\leftrightarrow -$). 
Since the transformation $U_{\bm{k}}$ is the same for both the bath modes and the emitter modes, the Chern numbers of the polariton bands will be the same as those of the original bath's bands. 
On the other hand, in the strong-coupling limit, $g\gg \abs{\Delta - \omega_j(\bm{k})}$, each bath's band gives rise to two strongly-hybridized polariton bands with dispersions $\lambda_{j,\pm}(\bm{k})\simeq [\Delta + \omega_j(\bm{k})]/2 \pm g$, and eigenmodes $\beta_{\bm{k},j,\pm} \simeq (\tilde\sigma^+_{\bm{k},j} \pm \tilde a^\dag_{\bm{k},j})/\sqrt{2}$. 
It is easy to see that the Chern number of these strongly-hybridized-polariton bands is the same as the Chern number of the free bath band that originates them. 
In fact, from Eq.~\eqref{eq:eigenvalues} it is clear that there are no band touchings in the spectrum of the whole system, provided the free bath spectrum is gapped, $\omega_i(\bm{k})\neq\omega_j(\bm{k})$ for all $i\neq j$. 
Thus, the Chern numbers of all the bands are constant and do not depend on the coupling strength $g$.

\begin{figure}
    \centering
    \includegraphics{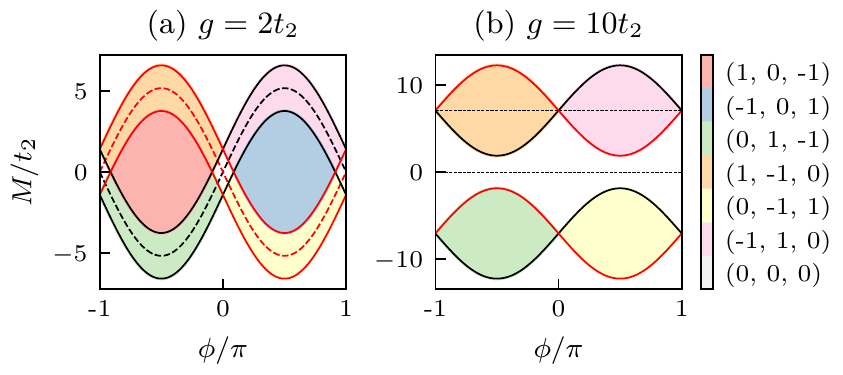}
    \caption{Topological phase diagram of the Haldane model with quantum emitters coupled to the $A$ sublattice in the weak (a) and strong (b) coupling regime. 
    Different phases are marked using different colors. 
    Each color corresponds to a different set of Chern numbers as indicated in the legend (from the lowest- to the highest-energy band). 
    The emitter frequency is tuned in all the cases to the middle of the inner bandgap $\Delta = 3t_2\cos(\phi)$. 
    The dashed red and black curves, with equation $M = \pm 3\sqrt{3}t_2\sin(\phi)$, correspond to the phase boundaries of the bare Haldane model. 
    In the presence of emitters, the boundaries are given by $M = \pm 3\sqrt{3}t_2\sin(\phi) \pm g/\sqrt{2}$ (solid red and black curves).}
    \label{fig:phasediags}
\end{figure}

A very different situation happens if we only couple emitters to a single sublattice. 
In the Markovian regime, it is easy to see that the Chern number of the single polariton band must be 0, since the sum of Chern numbers of all the bands must vanish \cite{Avron1983}, and in the limit $g\to 0$ the presence of emitters does not change the Chern numbers of the original bath's bands. 
On the other hand, the strongly-hybridized-polariton bands should also have a zero Chern number, since the corresponding eigenmodes are approximately given by $\beta^\dag_\pm \simeq(\sigma^\dag_{\bm{k}} \pm a^\dag_{\bm{k},1})/\sqrt{2}$ (assuming, w.l.o.g. that the emitters are coupled to the first sublattice). 
This behavior is exemplified in Fig.~\ref{fig:phasediags}, where we show two topological phase diagrams for a system of quantum emitters coupled to the $A$ sublattice of the Haldane model. 
For small coupling strengths the phase diagram is similar to that of the bare Haldane model, with new phases emerging near the original phase transition points. 
The main two lobes (red and blue regions) correspond to models where the center band is trivial (Markovian polariton band) and the bath's bands remain unaffected by the presence of emitters. 
As we increase $g$, the area of these two lobes is reduced, until they completely disappear for $g>3\sqrt{6}t_2$. 
In the limit $g\to\infty$ all the bands become trivial (fixing the values of the rest of parameters).
This can be understood noting that, in this limit, the emitters effectively decouple the two sublattices of the Haldane model.
As this case shows, the presence of emitters can also modify the topology of the guided modes in the bath.

We can argue intuitively that for strongly-hybridized polariton bands it is necessary to have a dense emitter superlattice to have non-trivial Chern numbers, since the photonic component of these bound states is very localized.
For sparse emitter supperlatices, topological polariton bands only occur in the Markovian regime.
Furthermore, for dense superlattices the bands of guided modes in the bath can become trivial in the strong-coupling limit, while for sufficiently sparse lattices, the topology of the bath's bands remains unaffected by the presence of emitters.

\section{Conclusion \label{sec:conclusion}}

To sum up, in this work we have explored the physics of polaritons appearing in systems of quantum emitters coupled to two-dimensional topological baths. 
In particular, we have focused on two paradigmatic examples of topological insulators: the Haldane model and the Harper-Hofstadter model.
In the Haldane model, we have shown how, in a certain parameter regime, the effective couplings resemble the ones expected for the honeycomb lattice, and for emitter detunings in the bandgap they break time-reversal symmetry.
A single emitter coupled to the Haldane model can seed a quasibound state, which gives rise to truly long-range interactions scaling like $\sim 1/r$, where $r$ is the distance between the emitters.
In constrast with the quasibound states found for a single emitter in other lattices (e.g., the honeycomb lattice), it is not a vacancy-like bound state, and therefore is not robust to changes in the light-matter coupling constant. 
For one-dimensional emitter arrangements (emitter line defects) coupled to the Haldane model in the Markovian regime, the gap of the effective polariton bands in the topological phase is orders of magnitude smaller than the one found in the trivial phase, and in the former case the polariton dynamics is chiral.
In the Harper-Hofstadter model, the Chern numbers of the surrounding bands dictate the non-monotonicity of the effective couplings, which give rise to a characteristic ballistic diffusion of polaritons with several wavefronts.
For two-dimensional emitter arrangements (emitter superlattices), we have shown how, in the case of one emitter coupled to every bath site, the resulting polariton bands always have the same Chern numbers of the original bath's bands.
For other geometries, however, it is possible to have different Chern numbers, and even change the topology of the bath's guided modes.

One aspect that we have not addressed in this work is the effect of disorder and imperfections in the emitter arrangements. 
For example, it could be interesting to consider systems where the emitters are coupled at random locations in the bath, and see whether the resulting amorphous effective models are topologically non-trivial \cite{Mitchell2018}.
Another interesting future research direction is to go beyond single-excitation physics.
Within the Markovian regime, the emitter dynamics can be described with an effective XX-type spin Hamiltonian. 
Thus, for an emitter line-defect in the Harper-Hofstadter model, the non-monotonic long-range couplings could result in multiple-component Luttinger-liquid phases.
In the Haldane model, the possibility of having complex, truly long-range couplings could result in interesting many-body phases that have not been described previously.

% It is interesting to see that polariton bands in emitter superlattices can also realize Chern numbers which are not present in the original free bath. 

\section*{Acknowledgements}
M.B. and J.I.C. acknowledge funding from the ERC Advanced Grant QUENOCOBA under the EU Horizon 2020 program (Grant
Agreement No. 742102).

\appendix

\section{Single-particle bound states \label{app:boundstates}}

In this section we review how to compute the single-particle bound states for a system of emitters coupled locally, i.e., each one to a single site of a $D$-dimensional bath, in the thermodynamic limit ($L\to\infty$). 
Their shapes and energies are intimately related to the properties of the level-shift operator (aka self-energy) $\Sigma(z)$ \cite{Cohen1992atom}. 
For such systems, the matrix elements of the level-shift operator in the basis of the single-excitation emitter subspace $\{\ket{n}\equiv \sigma_n^+\ket{\Omega}\}_{n=1}^{N}$, where $\ket{\Omega}\equiv \ket{g}^{\otimes N}\ket{\mathrm{vac}}$, are given by
\begin{equation}
\begin{split}
    \left[\Sigma(z)\right]_{mn} & \equiv \bra{m} \Sigma(z) \ket{n} \\
    & = g^2\left[\Phi(z,\bm{r}_m - \bm{r}_n)\right]_{\alpha_m,\alpha_n} \,,
\end{split}
\end{equation}
with (the integral is over the first Brillouin zone)
\begin{equation}
    \Phi(z,\bm{r}) = \int_\mathrm{BZ} \frac{d^D\bm{k}}{(2\pi)^D}\, e^{i\bm{k}\cdot\bm{r}} \left[z - \mathcal{H}_B(\bm{k})\right]^{-1} \,.
\end{equation}
With a slight abuse of notation, we can denote by $\Sigma(z)$ not the level-shift operator itself, but the above mentioned matrix.

Bound states are solutions to the eigenvalue equation $H\ket{\psi_\mathrm{BS}}=E_\mathrm{BS}\ket{\psi_\mathrm{BS}}$ with $E_\mathrm{BS}$ belonging to the discrete spectrum of $H$. Any state within the single-excitation subspace can be written as
\begin{equation}
    \ket{\psi_\mathrm{BS}} = \left[\sum_n c_{e,n}\sigma_n^+ + \sum_{\alpha,\,\bm{k}} c_{\alpha,\bm{k}} a_{\alpha,\bm{k}}^\dag\right]\ket{\Omega} \,.
\end{equation}
Denoting $\bm{c}_e \equiv \left(c_{e,1}, \dots, c_{e,N_e}\right)^T$, $\bm{c}_{\bm{k}} \equiv \left(c_{1,\bm{k}},\dots, c_{M,\bm{k}}\right)^T$, and $G_{\bm{k}}$ a $N_e\times M$ matrix with elements $\left(G_{\bm{k}}\right)_{nm}=g L^{-D/2} e^{i\bm{k}\bm{r}_n}\delta_{m,\alpha_n}$, the eigenvalue equation translates as
\begin{align}
    & \Delta \bm{c}_e + \sum_{\bm{k}} G_{\bm{k}} \bm{c}_{\bm{k}} = E_\mathrm{BS} \bm{c}_e \,, \label{eq:ce}\\
    & G_{\bm{k}}^\dag \bm{c}_e + \mathcal{H}_B(\bm{k}) \bm{c}_{\bm{k}}  = E_\mathrm{BS} \bm{c}_{\bm{k}} \,, \ \forall \bm{k}\in\mathrm{BZ}\,. \label{eq:ck}
\end{align}
From Eq.~\eqref{eq:ck},
\begin{equation}
    \bm{c}_{\bm{k}} = \left[E_\mathrm{BS} - \mathcal{H}_B(\bm{k})\right]^{-1} G_{\bm{k}}^\dag \bm{c}_e \,. \label{eq:ckk}
\end{equation}
Substituting back in Eq.~\eqref{eq:ce}, one can see that the amplitudes of the emitter excited states are nontrivial solutions of
\begin{equation}
    \left[\EBS - \Delta - \Sigma(\EBS)\right]\bm{c}_e = 0 \,, \label{eq:cee}
\end{equation}
therefore, the energies of all bound states (in the thermodynamic limit) are the solutions to the nonlinear equation
\begin{equation}
    \det \left[\EBS - \Delta - \Sigma(\EBS)\right] = 0 \,.
\end{equation}
For a given $\EBS$, the emitter amplitudes can be obtained solving Eq.~\eqref{eq:cee}, and with them, the photonic part can be computed using Eq.~\eqref{eq:ckk}. Importantly, Eq.~\eqref{eq:cee} does not fully determine $\bm{c}_e$. In order to do so, one has to use the normalization condition $\braket{\psi_\mathrm{BS}\vert\psi_\mathrm{BS}} = 1$, which implies
\begin{equation}
    \bm{c}_e^\dag \left[1 - \Sigma'(\EBS)\right]\bm{c}_e = 1 \,,
\end{equation}
where $\Sigma'(z) \equiv \partial_z \Sigma(z)$.
The bath amplitudes in position space can be obtained by an inverse Fourier transform of Eq.~\eqref{eq:ckk}, obtaining
\begin{equation}
    c_{\alpha,\bm{r}} = g\sum_{n} c_{e,n}\left[\Phi(\EBS,\bm{r} - \bm{r}_n)\right]_{\alpha,\alpha_n} \,. \label{eq:wavefun}
\end{equation}
Thus, knowing the self-energy, we can readily compute the photonic component of any bound state of a given energy.

\section{Approximate expressions for the self-energy in the Haldane model \label{app:selfenergyHaldane}}

The self-energy is essentially a Fourier transform of the kernel function 
\begin{equation}
    \mathcal K(\bm k) \equiv \frac{f_{\alpha_m\alpha_n}(z,\bm k)}{[z - h_0(\bm k)]^2 - h^2(\bm k)} \equiv \frac{f_{\alpha_m\alpha_n}(z,\bm k)}{d(z, \bm k)}\,. 
\end{equation}
Note that $d(z, \bm k) = \det [z - \mathcal H_B(\bm k)]$.
For $\abs{\bm r_{mn}}\gg 1$, the integrand is highly oscillatory, so it is the sharpest features of $\mathcal K$ that give non-negligible contributions to the integral. 
If $z = \Delta$ lies above or below the bands, in the regime $\abs{M},\abs{t_2} \ll \abs{t_1}$, the magnitude of the kernel is peaked around the $\Gamma$-point, $\bm \Gamma \equiv (0,0)$, which is actually an extremum of $d$. 
Thus, we can obtain an approximation to the self-energy expanding 
\begin{equation}
    d(\Delta, \bm \Gamma + \bm q)\simeq d(\Delta, \bm \Gamma) + \frac{1}{2} \bm q^T \,{\rm H}\, \bm q \,,
\end{equation}
where $\rm H$ denotes the Hessian of $d$ evaluated at the $\Gamma$-point,
which turns out to be of the form
\begin{equation}
    \rm H = 2c \begin{pmatrix}
    1 & -1/2 \\ -1/2 & 1
    \end{pmatrix} \,.
\end{equation}
With a linear transformation
\begin{equation}
    \bm p = \frac{1}{2}\begin{pmatrix}
    \sqrt{3} & -\sqrt{3} \\
    1        & 1
    \end{pmatrix} \bm q \equiv L \bm q \,,
\end{equation}
we can express
\begin{equation}
    \Sigma_{mn}(\Delta) \simeq \frac{g^2}{(2\pi)^2} \int_{\mathbb R^2} d^2\bm p\, \frac{a e^{i\bm p \cdot \bm r_{mn}'}}{b + cp^2} \,,
\end{equation}
where $\bm r'_{mn} = (L^{-1})^T \bm r_{mn}$, $p\equiv \norm{\bm p}$, and real constants
\begin{equation}
    \begin{split}
        a & = f_{\alpha_m\alpha_n}(\Delta,\bm \Gamma) \abs{\det L}^{-1} \\
        & = \frac{2}{\sqrt{3}} \times \begin{cases} 
        \Delta + 6 t_2 \cos(\phi) \pm M\,, & \text{if } \alpha_m = \alpha_n \\
        -3t_1 \,, & \text{if } \alpha_m \neq \alpha_n 
        \end{cases}
    \end{split} \label{eq:ctea}
\end{equation}
\begin{align}
    b & = d(\Delta,\bm \Gamma) =[\Delta + 6 t_2 \cos(\phi)]^2 - M^2 - 9 t_1^2 \,, \\
    c & = 2t_1^2 - 4t_2\cos(\phi)[\Delta + 6 t_2 \cos(\phi)] \,.
\end{align}
In Eq.~\eqref{eq:ctea}, the term $+M$ or $-M$ has to be chosen if $\alpha_m = A$ or $\alpha_m = B$, respectively.
Using polar coordinates, integrating the angular variable using the Jacobi-Anger expansion, we obtain
\begin{equation}
    \Sigma_{mn}(\Delta) \simeq \frac{g^2 a}{2\pi} \int_0^\infty dp\, \frac{p \J_0(p r'_{mn})}{b + c p^2} \,.
\end{equation}
Last, integrating the radial variable we obtain
\begin{equation}
    \Sigma_{mn}(\Delta) \simeq \frac{g^2 a}{2\pi c} \K_0\left(\sqrt{\frac{b}{c}} r'_{mn}\right) \,. \label{eq:selfe_approx}
\end{equation}
In these expressions, $\J_n$ and $\K_n$ denote the $n$th-order Bessel function of the first kind and modified Bessel function of the second kind, respectively.

When $\Delta$ lies in between the two bands, the sharpest features of the integrand are instead located at the K-points, $\pm \bm K\equiv \pm(1,-1)2\pi/3$.
Thus, following an analogous procedure for each of these points, we can approximate the self-energy by the sum of two different contributions $\Sigma_{mn}(\Delta)\simeq I_{mn}^+(\Delta) + I_{mn}^-(\Delta)$. 
When $\alpha_m = \alpha_n$, $I^\pm_{mn}(\Delta)$ has the same form as shown in the Eq.~\eqref{eq:selfe_approx}, with
\begin{equation}
\begin{split}
    & a_\pm = \frac{2 e^{\pm i \bm K \cdot \bm r_{mn}}}{\sqrt{3}} \\ 
    & \times \begin{cases}
    \Delta + M - 3 t_2[\shortcos(\phi) \mp \sqrt{3}\shortsin(\phi)] \,, & \text{if }\alpha_m = A  \\
    \Delta - M - 3 t_2[\shortcos(\phi) \pm \sqrt{3}\shortsin(\phi)] \,, & \text{if }\alpha_m = B \\
    \end{cases}
\end{split}
\end{equation}
\begin{align}
    b_\pm & = [\Delta - 3 t_2 \shortcos(\phi)]^2 - [M \pm 3\sqrt{3}t_2 \shortsin(\phi)]^2 \,, \label{eq:bpm}\\
    \begin{split}
    c_\pm & = -t_1^2 + 6 t_2^2 [1 - 2\shortcos(2\phi)] \\
    & \qquad + 2 t_2 [\Delta \shortcos(\phi) \pm \sqrt{3} M \shortsin(\phi)] \,,
    \end{split} \label{eq:cpm}
\end{align}
where we have used the shorthand notation $\shortcos(\phi)\equiv \cos(\phi)$ and $\shortsin(\phi)\equiv \sin(\phi)$.
For $\alpha_m\neq\alpha_n$, the numerator in the kernel vanishes at zeroth order, $f_{\alpha_m\alpha_n}(\Delta,\pm\bm K) = 0$, so we have to consider the next order to obtain a good approximation. 
Doing so, using polar coordinates and integrating the angular variable, we obtain 
\begin{equation}
\begin{split}
    I^\pm_{mn}(\Delta) & = \frac{g^2 a_\pm}{2\pi} \int_0^\infty dp\, \frac{p^2 \J_1(p r'_{mn})}{b_\pm + c_\pm p^2} \\
    & = \frac{g^2 a_\pm}{2\pi c_\pm} \sqrt{\frac{b_\pm}{c_\pm}} \K_1\left(\sqrt{\frac{b_\pm}{c_\pm}} r'_{mn}\right) \,,
\end{split}
\end{equation}
where $b_\pm$ and $c_\pm$ are the same as in Eqs.~\eqref{eq:bpm} and \eqref{eq:cpm}, and 
% \begin{equation}
%     a_\pm = \frac{2 t_1 e^{\pm i\bm K\cdot \bm r_{mn}}}{\sqrt{3}}\times \begin{cases}
%     e^{\mp i (\theta_{mn} - \pi/2)} \,, & \text{if } \alpha_m = A\\
%     e^{\pm i (\theta_{mn} + \pi/2)}  \,, & \text{if } \alpha_m = B
%     \end{cases} \,.
% \end{equation}
% Here, $\theta_{mn}$ is the angle between $\bm r'_{mn}$ and the horizontal axis. 
% Note that with a simple gauge transformation, multiplying all spin operators within the same sublattice by the same phase factor, we can add an arbitrary \emph{constant} phase to $J_{mn}$ for the case $\alpha_m = A$ and $\alpha_n = B$ (the opposite phase for $\alpha_m = B$ and $\alpha_n = A$). 
% Thus, we can actually consider 
\begin{equation}
    a_\pm = \frac{2 t_1 e^{\pm i\bm K\cdot \bm r_{mn}}}{\sqrt{3}}\times \begin{cases}
    e^{\mp i (\theta_{mn} - \pi/2)} \,, & \text{if } \alpha_m = A\\
    e^{\pm i (\theta_{mn} + \pi/2)}  \,, & \text{if } \alpha_m = B
    \end{cases} \,.
\end{equation}
Here, $\theta_{mn}$ is the angle between $\bm r'_{mn}$ and the horizontal axis. 

\section{One-dimensional self-energies} \label{app:1D}

When viewing 2D systems with an emitter line-defect as a collection of 1D systems, the self-energy for each one of the 1D systems can be obtained using Eq.~\eqref{eq:phi}, integrating only the quasimomentum perpendicular to the emitter line-defect.

\subsection{Harper-Hofstadter model} \label{app:1DHarperHofstadter}

In this case, $H_B^\mathrm{1D}(k)$ corresponds to a chain with nearest-neighbor hopping of amplitude $J$, and on-site chemical potentials $\{\mu_n=-2J\cos(k + n\phi):n\in\mathbb{Z}\}$, 
\begin{equation}
    H_B^{\rm 1D} = -J\sum_n \left(c^\dag_{n+1} c_n + c^\dag_n c_{n+1}\right) + \sum_n \mu_n c^\dag_n c_n \,,
    \label{eq:periodicbath1D}
\end{equation}
where we assume that the bosonic creation and annihilation operators $\{c^\dag_n, c_n\}_{n\in\mathbb{Z}}$ depend implicitly on the parallel quasimomentum $k$.
Since $\phi = 2\pi p/q$ with co-prime integers $p$ and $q$, the chemical potentials are periodic with period $q$ (there are $q$ sites per unit cell).
We may assume that the emitter is coupled at $n = 1$, since a redefinition of the quasimomentum $k_x\to k_x + m \phi$ (equivalent to choosing a different gauge for the magnetic vector potential) corresponds to a cyclic permutation of the chemical potentials $\mu_n\to\mu_{n+m}$. 
In addition, the system is invariant under the combined action of time-reversal and space inversion about the emitter position, $\mu_n(k-\phi) = \mu_{q - n + 2}(-k-\phi)$.

The matrix $\left[z-\mathcal{H}_B(k_y)\right]/J$ is of the form 
\begin{equation}
    A = \begin{pmatrix}
    a_1 & 1 & & & & e^{-ik_y}\\
    1 & a_2 & 1 & & & \\
    & 1 & \ddots & \ddots & & \\
    & & \ddots & & &\\
    & & & & & 1\\
    e^{ik_y} & & & & 1 & a_q\end{pmatrix} \,,
\end{equation}
with $a_n = (z - \mu_n)/J$. To compute the self-energy, we have to compute its inverse, whose matrix elements are given by $(A^{-1})_{mn} = C_{nm}/\det A$. Here $C$ is the cofactor matrix associated to $A$. Using Laplace's formula to compute determinants, we find (for $q\geq 3$ and $2<p<q$)
\begin{align}
    C_{11} & = K(a_2,\dots,a_q) \,, \label{eq:cofac1} \\
    C_{12} & = -\left[K(a_3,\dots,a_q) + (-1)^q e^{ik_y}\right] \,,
    \\
    \begin{split}
    C_{1p} & = (-1)^{p+1}\big[K(a_{p+1},\dots,a_q) \\
    & \qquad\qquad\qquad + (-1)^q e^{ik_y}K(a_2,\dots,a_{p - 1})\big] \,,
    \end{split}
    \\
    C_{1q} & = (-1)^{q+1}\left[1 + (-1)^q e^{ik_y}K(a_2,\dots,a_{q - 1}) \right] \,, \label{eq:cofacq} \\
    \begin{split}
    \det A & = K(a_1,\dots,a_q) - K(a_2,\dots,a_{q-1}) \\ 
    & \qquad + (-1)^{q+1}2\cos k_y \,, 
    \end{split}
\end{align}
where we have expressed the different quantities using the continuant $K$, defined as
\begin{equation}
    K(x_1,\dots,x_r) \equiv \det \begin{pmatrix}
    x_1 & 1 & & & & \\
    1 & x_2 & 1 & & & \\
    & 1 & \ddots & 
    \ddots & & \\
    & & \ddots & & &\\
    & & & & & 1\\
    & & & & 1 & x_r
    \end{pmatrix} \,.
\end{equation}
Note that this continuant satisfies the recurrence relation: 
\begin{equation}
    K_n = x_n K_{n-1} - K_{n-2} \,,\quad 1\leq n\leq r\,,
\end{equation}
where $K_n\equiv K(x_1, \dots, x_n)$, and we set $K_0=1$, $K_{-1}=0$.

Finally, the integral giving the emitter self-energy is straightforward to compute using residue integration. For example:
\begin{widetext}
\begin{align}
    \left[\Phi(z, y)\right]_{11} & = \frac{1}{2\pi J}\int_{-\pi}^\pi dk_y\, \frac{K(a_2,\dots,a_q)e^{ik_y y}}{K(a_1,\dots,a_q)-K(a_2,\dots,a_{q-1})+(-1)^{q+1}2\cos k_y} \label{eq:step1} \\
    & = \frac{(-1)^{q+1}K(a_2,\dots,a_q)}{2\pi i J}\oint dw\,\frac{w^{\abs{y}}}{w^2 + bw + 1} \label{eq:step2} \\
    & = \frac{(-1)^{q+1}K(a_2,\dots,a_q)}{J\sqrt{b^2 - 4}}\left[w_+^{\abs{y}}\Theta(1 - \abs{w_+}) - w_-^{\abs{y}}\Theta(1 - \abs{w_-})\right] \,. \label{eq:step3}
\end{align}
\end{widetext}
To go from Eq.~\eqref{eq:step1} to Eq.~\eqref{eq:step2} we perform a change of variable $w=\exp[ik_y\sign(y)]$, such that the integral becomes a contour integral in the complex plane along the unit circumference (anticlockwise). Last, in Eq.~\eqref{eq:step3} we express the result in terms of the poles $w_\pm = (-b \pm \sqrt{b^2 - 4})/2$, which are the roots of the second-order polynomial $p(w) = w^2 + bw + 1$, with
\begin{equation}
    b = (-1)^{q+1}\left[K(a_1,\dots,a_q)-K(a_2,\dots,a_{q-1})\right] \,; 
\end{equation}
$\Theta$ denotes Heaviside's step function. Since $p$ is a palindromic polynomial (its coefficients form a palindrome), its roots are one the inverse of the other, $w_+w_- = 1$. As a consequence, only one of the poles contributes to the integral in general, except when $\abs{w_+}=\abs{w_-}=1$, which occurs whenever $z$ belongs to the spectrum of $H^\mathrm{1D}_B$. In that case, the integral in Eq.~\eqref{eq:step2} is not well defined, and the self-energy is discontinuous. For real $z$, $b$ is real, and if $z$ is not in the spectrum of $H^\mathrm{1D}_B$, then the poles $w_\pm$ are also real. In this case, the pole inside the unit circumference is $w_\mathrm{in} = w_{\sign(b)}$, and Eq.~\eqref{eq:step3} can be simplified as 
\begin{equation}
    \left[\Phi(z,y)\right]_{11} = \frac{\sign(b)(-1)^{q+1}K(a_2,\dots,a_q)}{J\sqrt{b^2 - 4}}w_\mathrm{in}^{\abs{y}}\,. \label{eq:phi11}
\end{equation}
The self-energy appearing in the pole equation, Eq.~\eqref{eq:pole1d}, used to compute the polariton band is simply $\Sigma(\omega_\mathrm{BS}) = g^2 \left[\Phi(\omega_\mathrm{BS},0)\right]_{11}$. It depends implicitly on the quasimomentum $k_x$ through the on-site energies $\{a_n\}_{n=1}^q$. The zeroes of this function, which correspond to the vacancy-like bound states, are the zeroes of the continuant $K(a_2,\dots,a_q)$.

Once we know the energy of a bound state, we can compute analytically its wavefunction using Eq.~\eqref{eq:wavefun}. For this, we need to compute $\left[\Phi(z,y)\right]_{n1}$, for real $z$ not in the spectrum of $H^\mathrm{1D}_B$. From Eqs.~\eqref{eq:cofac1} to \eqref{eq:cofacq}, we realize that the cofactors have the form $C_{1n}=F_n + G_n e^{ik_y}$. Thus,
\begin{equation}
    \left[\Phi(z,y)\right]_{n1} = \frac{\sign(b)(-1)^{q+1}\left[F_n + G_n w_\mathrm{in}^{\sign(y)}\right]}{J\sqrt{b^2 - 4}}w_\mathrm{in}^{\abs{y}}\,. \label{eq:phin1}
\end{equation}

Eqs.~\eqref{eq:phi11} and \eqref{eq:phin1} allow us to extract the following conclusions regarding the shape of the bound states:
\begin{enumerate}
    \item On the sublattice to which the emitter is coupled, the amplitudes are symmetric with respect to the emitter position. This is not the case for the other sublattices.
    \item In general, the bound state is localized exponentially around the emitter position, with the same decay length in all sublattices. It may decay monotonically or alternate sign, depending on the sign of $w_\mathrm{in}$.
    \item For a vacancy-like bound state, the amplitude of the wavefunction on the sublattice to which the emitter is coupled vanishes everywhere. 
\end{enumerate}

Except in the case of accidental degeneracies, the vacancy-like bound states are completely localized on one side of the emitter. 
This can be understood, since the vacancy modes of the system are the eigenstates of the two semi-infinite chains in which the emitter divides the bath. As each chain is characterized by a different order of the on-site potentials, the vacancy modes will not be degenerate in general.

We remark that the conclusions drawn here are valid for arbitrary 1D systems of the form shown in Eq.~\eqref{eq:periodicbath1D} and not just for the emitter-line defects in the Harper-Hofstadter model.
That is, Eqs.~\eqref{eq:phi11} and \eqref{eq:phin1} are valid regardless the specific values of the chemical potentials, as long as they are $q$-periodic, $\mu_n = \mu_{n+q}$.

\subsection{Haldane model} \label{app:1DHaldane}

\begin{figure}[!h]
    \centering
    \includegraphics{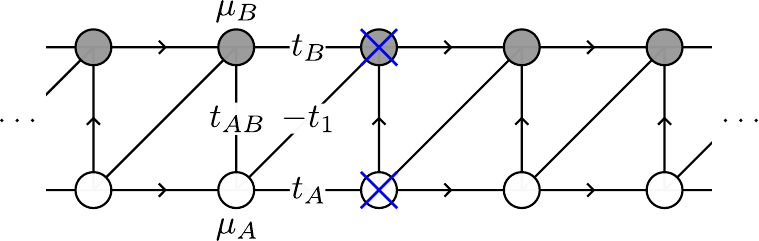}
    \caption{(a) Schematics of the 1D model relevant for the study of the bound states in the Haldane model. The on-site potentials depend on the original model's parameters as
    $\mu_A = -2 t_2 \cos(k_x + \phi) + M$, and $\mu_B(\phi,M)=\mu_A(-\phi,-M)$; the hoppings, which may be complex, are given by $t_{AA} = -t_2 e^{i\phi} -t_2 e^{i(k_x-\phi)}$, $t_{BB} = t_{AA}(-\phi)$, $t_{AB}=-t_1$, $t_{BA}=-t_1(1+e^{ik_x})$. 
    The arrows indicate the hopping direction, with the opposite direction having the complex-conjugated hopping amplitude.} 
    \label{fig:drawing_haldane_1D}
\end{figure}

In this case, $H_B^\mathrm{1D}(k)$ corresponds to a ladder with complex hopping amplitudes, such as the one shown in Fig.~\ref{fig:drawing_haldane_1D}. 

Instead of focusing on the Haldane model, in this section we describe a general procedure to compute the self-energy, valid for any two-band, one-dimensional bath with finite hopping range $R$. 
The momentum-resolved Hamiltonian for such models can always be expressed in terms of the Pauli matrices $\{\sigma^\nu\}_{\nu=x,y,z}$ and the identity matrix $\sigma^0$, as $\mathcal{H}_B(k) = \sum_\alpha h_\alpha(k)\sigma^\alpha$. 
The functions $h_\alpha$ are finite Laurent series of the variable $w = e^{ik}$, i.e., they are functions of the form 
\begin{equation}
    h_\alpha(w) = a_0 w^{-R} + a_1 w^{-R + 1} + \dots + a_{2R} w^R\,,
\end{equation}
so they are completely determined by the vector of coefficients $[a_0, \dots, a_{2R}]$, which are themselves simple linear combinations of the hopping amplitudes of the model. We define the polynomials $p_\alpha(w) = w^R h_\alpha(w)$, which have the same vector of coefficients in the standard polynomial basis. Hermiticity of $\mathcal{H}_B$ then implies that these polynomials are self-reciprocal, i.e., their coefficients satisfy $a_n^* = a_{2R - n}$, for $0\leq n\leq R$.

For a 2-by-2 matrix, the inverse $\left(z-\mathcal{H}_B\right)^{-1} = C^T/\det\left(z-\mathcal{H}_B\right)$ can be computed straightforwardly,
\begin{align}
    & \det\left(z-\mathcal{H}_B\right) = (z-h_0)^2 - h_x^2 - h_y^2 - h_z^2 \\
    & C^T =  \begin{pmatrix}
    z-h_0+h_z & h_x - ih_y \\
    h_x+ih_y & z-h_0-h_z
    \end{pmatrix} \,.
\end{align}
Thus, doing a change of variable $w=\exp[\sign(y) i k]$, we can express the self-energy as an integral of a rational function over the unit cricumference in the complex plane, 
\begin{equation}
    \left[\Phi(z,y)\right]_{\alpha\beta} = \frac{1}{2\pi i}\oint dw\, w^{\abs{y} + R -1} \frac{P_{\alpha\beta}(w)}{Q(w)} \,.
\end{equation}
For $y \geq 0$,
\begin{equation}
    Q(w) = \left[zw^R-p_0(w)\right]^2 - p_x(w) - p_y(w) - p_z(w) \,,
\end{equation}
while the numerator is, depending on the sublattice indices, given by
\begin{align}
    & P_{AA}(w) = zw^R - p_0(w) + p_z(w) \,, \\
    & P_{BB}(w) = zw^R - p_0(w) - p_z(w) \,, \\
    & P_{AB}(w) = p_x(w) - i p_y(w) \,, \\
    & P_{BA}(w) = p_x(w) + i p_y(w) \,.
\end{align}
For $y < 0$, the same expressions for both $P_{\alpha\beta}$ and $Q$ can be used, replacing $p_\alpha$ by $\overline{p}_\alpha(w)=w^Rh_\alpha(w^{-1})$, which is the same polynomial as $p_\alpha$ with the coefficients in reverse order. While $Q$ is not palindromic (its coefficients are complex in general), for real $z$ it is self-reciprocal. This implies that its roots come in pairs of the form: $w_j$, $1/w_j^*$. So, when $z$ is real and it lies outside the band regions, there will be at most $R$ poles $\{w_{\mathrm{in},j}\}_{j=1}^S$, $S\leq R$, inside the unit circumference that contribute to the integral. Therefore, we can express
\begin{equation}
    \left[\Phi(z,y)\right]_{\alpha\beta} = \sum_{j} \left(w_{\mathrm{in},j}\right)^{\abs{y}+R-1} \res\left(\frac{P_{\alpha\beta}}{Q},w_{\mathrm{in},j}\right) \,.
\end{equation}
As the hopping range increases, computing the poles and residues manually can become challenging. However, this task can be carried out using well-established numerical algorithms \cite{Mahoney1983}.

The self energy matrix appearing in the ``pole equation'', Eq.~\eqref{eq:pole1d2emitter}, is given by $\Sigma(\omega_\mathrm{BS}) = g^2\Phi(\omega_\mathrm{BS},0)$.

\section{Strong light-matter coupling \label{app:strong_light-matter}}

When the light-matter interaction is the leading energy scale, $g\gg \Delta, J$, it is convenient to work in a basis that diagonalizes $H_I$. So, we consider the hybrid modes $\beta_{n,\pm}\equiv \left(\sigma^-_n \pm a_{\bm{r}_n,\alpha_n}\right)/\sqrt{2}$, with which we can write
\begin{equation}
    H_I = g \sum_n(\beta^\dag_{n,+}\beta_{n,+} - \beta^\dag_{n,-}\beta_{n,-})\,.
\end{equation}
$H_S$ and $H_B^\bullet$ now couple these hybrid modes between themselves and with the rest of bath modes. For example, for the 1D system corresponding to a line-defect in the Harper-Hofstadter model, we have
\begin{align}
    H_S & = \frac{\Delta}{2} (\beta^\dag_++\beta^\dag_-)(\beta_++\beta_-) \,,\\
    \begin{split}
      H_B^\bullet & = \frac{\mu_1}{2}(\beta^\dag_+-\beta^\dag_-)(\beta_+-\beta_-) \\
      & \quad - \frac{J}{\sqrt{2}}\sum_{j=0,2}\left[(\beta^\dag_+-\beta^\dag_-) b_j + b^\dag_j(\beta_+-\beta_-)\right] \,.
    \end{split}
\end{align}
Here we have assumed that the emitter is coupled to a site with chemical potential $\mu_1$.
The total Hamiltonian can be represented graphically as shown in Fig.~\ref{fig:strongly-hybr_HH} below.
\begin{figure}[!htb]
    \centering
    \includegraphics{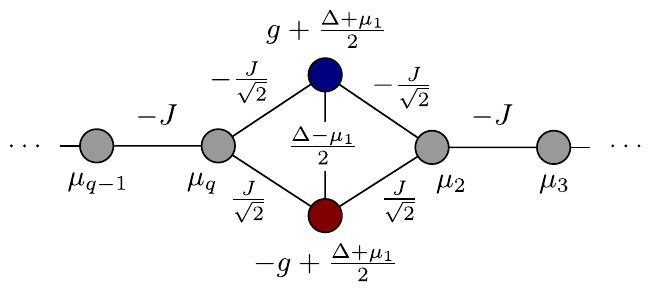}
    \caption{Schematic representation of the total Hamiltonian as a graph, written in the basis $\beta_+$ (blue), $\beta_-$ (red) and $\{b_n\}_{n\neq 1}$ (gray). The hopping strength is indicated next to each link, while on-site energies are indicated above or below each site.}
    \label{fig:strongly-hybr_HH}
\end{figure}

Using standard perturbative methods \cite{Bravyi2011}, we can obtain effective Hamiltonians for these strongly-hybridized polariton modes. Within the single-particle sector, the Hamiltonian can be written as
\begin{equation}
    H = \sum_{m,\,n} J_{mn}\sigma_{mn}\,,\quad \sigma_{mn}\equiv \ket{m}\!\bra{n} \,,
\end{equation}
where $\{\ket{n}\}_{n\geq 0}$ is a basis of the single-particle subspace of states. If $\abs{J_{00}-J_{nn}}\gg \abs{J_{0n}}$ for $n>0$, then, a unitary transformation
\begin{equation}
    \tilde H \equiv e^{-S} H e^S \simeq H + [H, S] + \frac{1}{2!}\left[[H, S],S\right] + \cdots
\end{equation}
with
\begin{equation}
    S = \sum_n \frac{1}{J_{nn}-J_{00}}\left(J_{0n}\sigma_{0n} - J_{n0}\sigma_{n0}\right)\,,
\end{equation}
can remove all zeroth-order off-diagonal terms coupling the state $\ket{\tilde{0}}$ with the rest of states in the new basis, $\{\ket{\tilde{n}}\equiv e^{-S}\ket{n}\}_n$, leaving only terms $\propto (J_{00} - J_{nn})^{-\alpha}$ with $\alpha\geq 1$. Thus, to first order in perturbation series, the dispersion relation of the strongly-hybridized polaritons appearing in the model shown in Fig.~\ref{fig:strongly-hybr_HH} can be computed as
\begin{equation}
\begin{split}
    \omega_\pm & = \pm g + \frac{\Delta + \mu_1}{2}  \pm \frac{(\Delta-\mu_1)^2}{8 g} \\ & \qquad + \sum_{\alpha=2,q}\frac{J^2}{\pm 2g + \Delta + \mu_1 - 2\mu_\alpha} \,.
\end{split}
\label{eq:approx_dispersion}
\end{equation}

While in the weak coupling limit the emitter dynamics can be modelled with an effective single-band model, in the strongly-interacting regime the effective emitter Hamiltonian has two bands, one for the ``symmetric'' and another for the ``antisymmetric'' strongly-hybridized polaritons, that appear above and below the bath's energy bands, respectively. 
This symmetry with respect to the exchange of the excitation between the emitters and the bath is an approximate conserved quantity. 
From Eq.~\eqref{eq:approx_dispersion} we can see that, to leading order, the dispersion of each kind of polariton has the usual cosine shape of a 1D tight-binding model with nearest-neighbor hopping of strength $J/2$.
As a consequence, an initially localized excitation has the usual ballistic diffusion along the emitter line with a single wavefront.

A similar analysis can be carried out for the Haldane model. 
In this model, there are four polariton bands that appear above and below the bath's bands. 
Interestingly, while the minimal gap of the polaritons' spectrum still changes depending on the topology of the underlying bath, it remains more or less of the same order. 
In addition, the average chiral group velocity is much smaller than in the weak-coupling regime (relative to the effective hopping). % , see Fig.~\ref{fig:haldane_strong}.

% \begin{figure}
%     \centering
%     \includegraphics{haldane_strong.pdf}
%     \caption{$\mathrm{Gap}\equiv \min_{k_x,\,\alpha,\,\beta}\abs{\omega_{\mathrm{BS},\alpha}(k_x) - \omega_{\mathrm{BS},\beta}(k_x)}$, and average chiral group velocity for a system with the same parameters as in Fig.~\ref{fig:haldane_1D}(b), but with a strong light-matter coupling $g=5.5t_1$.}
%     \label{fig:haldane_strong}
% \end{figure}

\bibliography{quantum_emitters.bib}

\end{document}